\documentclass[aps,superscriptaddress, showpacs,preprintnumbers,  nofootinbibt,twocolumn]{revtex4}
\usepackage{amssymb}
\usepackage{graphicx}

\def\be{\begin{equation}}
\def\ee{\end{equation}}
\def\bea{\begin{eqnarray}}
\def\eea{\end{eqnarray}}

\begin{document}

\title{Irreversible thermodynamic description of dark matter and radiation creation during inflationary reheating}

\author{Juntong Su}
\email{juntong.su@outlook.com}
\affiliation{School of Physics, Sun Yat-Sen University, Guangzhou 510275, P. R. China}
\affiliation{Yat Sen School, Sun Yat-Sen University, Guangzhou 510275, P. R. China}
\author{Tiberiu Harko}
\email{t.harko@ucl.ac.uk}
\affiliation{Department of Physics, Babes-Bolyai University, Kogalniceanu Street, Cluj-Napoca 400084, Romania}
\affiliation{Department of Mathematics, University College London, Gower Street, London
WC1E 6BT, United Kingdom}
\author{Shi-Dong Liang}
\email{stslsd@mail.sysu.edu.cn}
\affiliation{School of Physics, Sun Yat-Sen University, Guangzhou 510275, P. R. China}
\affiliation{State Key Laboratory of Optoelectronic Material and Technology, and
Guangdong Province Key Laboratory of Display Material and Technology, Guangzhou, P. R. China}

\date{\today }

\begin{abstract}
We investigate the matter creation processes during the reheating period at the end of inflation in the early Universe, by using the irreversible thermodynamic of open systems. The matter content of the Universe is assumed to consist of the inflationary scalar field, which, through its decay, generates relativistic matter, and pressureless dark matter, respectively. At the early stages of reheating the inflationary scalar field transfers its energy to the newly created matter particles, with the field energy decreasing to near zero. The general equations governing the irreversible matter creation during reheating are obtained by combining the thermodynamics description of the matter creation and the gravitational field equations. A dimensionless form of the general system of the reheating equations is also introduced. The role of the different inflationary scalar field potentials is analyzed by using analytical and numerical methods, and the evolution of the matter and scalar field densities, as well as of the cosmological parameters during reheating,  are obtained.  Typically, the values of the energy densities of relativistic matter and dark matter reach their maximum when the Universe is reheated up to the reheating temperature, which is determined for each case, as a function of the scalar field decay width, the scalar field particle mass, and of the cosmological parameters. An interesting result is that particle production leads to the acceleration of the Universe during the reheating phase, with the deceleration parameter showing a complex dynamics. Once the energy density of the scalar field becomes negligible with respect to the matter densities,  the expansion of the Universe decelerates, and inflation has a graceful exit after reheating.


\end{abstract}

\maketitle

\section{Introduction}

The inflationary paradigm is one of the corner stones of present day cosmology. Inflation was first proposed in \cite{guth1981inflationary} to solve the spatial flatness, the horizon and the monopole problems of the early Universe. The main theoretical ingredients in inflation are a scalar field $\phi$ with self-interaction potential $V(\phi)$, and its energy density $\rho _{\phi}=\dot{\phi }^2/2+V(\phi)$, and pressure $p_{\phi}=\dot{\phi}^2/2-V(\phi)$, respectively \cite{B3, B31, B1, B2}. In the initial approach the inflation potential arrives at a local minimum at $\phi=0$ through supercooling from a phase transition, and then the Universe expands exponentially.  However, in this scenario, called "old inflation", there is no graceful exit to the inflationary stage. To solve this problem the "new inflation" model was proposed in \cite{linde1982new, linde1982coleman, linde1982scalar,albrecht1982cosmology}. In this model the inflationary scalar field is initially at its local maximum $\phi=0$. The potential is required to be very flat near the minimum at $\phi\neq 0$, so that the field rolls slowly down, without changing much before the Universe expands exponentially.   The "flatness" requirement is not easy to obtain, and another problem of this scenario is that the inflation has to begin very late and last a long time, or it may never happen in some cases \cite{linde2000inflationary}. By extending the new inflation model, the chaotic inflation theory was proposed \cite{linde1983chaotic}, which solved the problems above. The model does not require a thermal equilibrium state of the early Universe, and various forms of the scalar field potentials are acceptable, so that a large number  potential forms can be used in this scenario. Later a subtler theory of hybrid inflation with two scalar fields in the model was proposed \cite{linde1994hybrid}. The potential takes the form of $V(\sigma,\phi)=(M^2-\lambda\sigma^2)/(4\lambda)+m^2\phi^2/2+g^2\phi^2\sigma^2/2$, with $M$, $m$, $\lambda $, $\sigma $ constants. One scalar field accounts for the exponential growth of the Universe, and the other one is responsible for the graceful exit of inflation.

Presently, very precise observations of the Cosmic Microwave Background (CMB) radiation give us the opportunity to test the fundamental predictions of inflation on primordial fluctuations, such as scale independence and Gaussianity \cite{Pl1,Pl2,Pl3,Pl4,Pl5}. One can calculate cosmological parameters from CMB fluctuations, and use them as constraints of the inflationary models. The slow-roll parameters $\epsilon$ and $\eta$ can be calculated directly from the potentials of the inflationary scalar field, so that these parameters can be observed, and the corresponding inflationary models can be tested \cite{liddle1992cobe}. The slow-roll parameters and the inflationary parameters like scalar spectral index $n_s$ and tensor-to-scalar ratio $r$ of several forms of the inflationary potentials were investigated in \cite{dodelson1997cosmic}. Further investigations  of the parameters $n_s$ and $r$ can be found in \cite{kinney1998constraining, kinney2000new, kinney2008latest}.

The prediction of the statistical isotropy, claiming that inflation removes the classical anisotropy of the Universe has been a central prediction of inflationary scenarios, also strongly supported  by the cosmic no-hair conjecture. However, recently several observations of the large scale structure of the Universe have raised the possibility that the principles of homogeneity and isotropy may not be valid at all scales, and that the presence of an intrinsic large scale anisotropy in the Universe cannot be ruled out {\it a priori} \cite{An}.  Even that generally the recent Planck observations tend to confirm the fundamental principles of the inflationary paradigm, the CMB data seem to show the existence of some tension between the models and the observations. For example, the combined Planck and  WMAP polarization data show that the index of the power spectrum is  $n_s = 0.9603\pm 0.0073$ \cite{Pl2,Pl3}, at the pivot scale $k_0 = 0.05$ Mpc$^{-1}$, a result which definitely rules out the exact scale-invariance ($n_s = 1$), as predicted by some inflationary models \cite{Haw,Muk},  at more than $5\sigma$. Moreover, this case can also be ruled out based on some fundamental theoretical considerations,  since it is connected with a de Sitter phase, and, more importantly,  it does not lead to a consistent inflationary model, mainly due to the graceful exit problem. On the other hand the joint constraints on $r$ (tensor-to-scalar ratio) and $n_s$ can put strong restrictions on the inﬂationary scenarios. For example, inﬂationary models with power-law potentials of the form  $\phi ^4$ cannot provide an acceptable number of e-folds (of the order of  50-60) in the restricted space  $r-n_s$ at a $2\sigma $ level \cite{Pl3}. In actuality, the theories of inflation have not yet reached a general consensus, because not only the overall inflationary picture, but also the details of the theoretical models are based on physics beyond the Standard Model of particle physics.

During inflation, the exponential growth of the Universe lead to a homogeneous, isotropic and empty universe. Elementary particles are believed to be created in the period of reheating at the end of inflation, defrosting the Universe, and transferring energy from the inflationary scalar field to matter. Reheating was suggested along with the new inflationary scenario \cite{linde1982new}, and further developed in \cite{Alb, Kof, kofman1997towards}. When the inflationary scalar field reaches its minimum after the expansion of the Universe, it oscillates around the minimum of the potential, and decays into Standard Model particles.  The newly created particles interact with each other, and reach a thermal equilibrium state of temperature $T$.
However, quantum field theoretical investigations indicate that the reheating period is characterised by complicated nonequilibrium processes, the main characteristic of which are initial, violent particle production via parametric resonance and inflaton decay (‘preheating’) with a highly nonequilibrium distribution of the produced particles, subsequently relaxing to an equilibrium state. Such explosive particle production is due to parametric amplification of quantum fluctuations for the nonbroken symmetry case (appropriate for chaotic inflation) or spinodal instabilities in the broken
symmetry phase \cite{Pre1,Pre2,Pre3,Pre4}.

The reheating temperature $T_{\text{reh}}$ is thought to be the maximum temperature of the radiation dominated Universe after reheating. However, it  was argued that the reheating temperature is not necessarily the maximum temperature of the Universe after reheating \cite{scherrer1985decaying}. In other words, reheating might be just one of the stages of the matter creation, and other substance with larger freeze-out temperature may be created after inflation through other process \cite{giudice2001largest}.

In the study of reheating,  the phenomenological approach introduced in \cite{Alb} has been widely investigated. This approach is based on the introduction of a suitable loss term in the scalar field equation, which also appears as a source term for the energy density of the newly created matter fluid. It is this source term that, if chosen appropriately, is responsible for the reheating process that follows adiabatic supercooling during the de Sitter phase. In this simple model  the corresponding dynamics is considered within a two component model. The first
component is a scalar field, with energy density and pressure $\rho _{\phi}$ and $p_{\phi}$, respectively, and with energy momentum tensor
$^{(\phi)}T_{\mu }^{\nu}=\left(\rho _{\phi}+p_{\phi}\right)u_{\mu}u^{\nu}-p_{\phi}\delta _{\mu}^{\nu}$.
The second component is represented by normal matter, with energy density and pressure $\rho _m$ and $p_m$, and with energy-momentum tensor
$^{(m)}T_{\mu }^{\nu}=\left(\rho _{m}+p_{m}\right)u_{\mu}u^{\nu}-p_{m}\delta _{\mu}^{\nu}$. In the above equations for simplicity we assume that both cosmological fluid components share the same four velocity $u_{\mu}$, normalized as $u_{\mu}u^{\mu}=1$. Einstein's field equations imply the relation $\nabla _{\mu}\left(^{(\phi)}T_{\mu }^{\nu}+^{(m)}T_{\mu }^{\nu}\right)=0$, which in the case of a flat geometry reduces to
\be\label{1}
\dot{\phi}\ddot{\phi}+3H\dot{\phi}^2+V'(\phi)\dot{\phi}+\dot{\rho }_m+3H\left(\rho _m+p_m\right)=0,
\ee
where $H=\dot{a}/a$ is the Hubble function, and $a$ is the scale factor of the Universe. In order to describe the transition process between the scalar field and the matter
component it is convenient to introduce phenomenologically  a ‘friction term’ $\Gamma $,  describing the decay
of the scalar field component, and acting as a source term for the matter fluid. Hence, Eq.~(\ref{1})  is assumed to decompose into two separate equations \be\label{phen}
\ddot{\phi}+3H\dot{\phi}+V'(\phi)+\Gamma \dot{\phi}=0,
\ee
\be
\dot{\rho }_m+3H\left(\rho _m+p_m\right)-\Gamma \dot{\phi}^2=0.
\ee

In the framework of the braneworld models, in which our Universe is a 3-brane embedded in a five-dimensional bulk, the reheating era after the inflationary period was analyzed \cite{Re11},  by assuming the possibility of brane-bulk energy exchange. The inflaton field was assumed to decay into normal matter only, while the dark matter is injected into the brane from the bulk.  The observational constraint of an approximately constant ratio of the dark and the baryonic matter requires that the dark matter must be non-relativistic (cold). The model predicts a reheating temperature of the order of $3 \times 10^6$ GeV, and brane tension of the order of $10^{25}$ GeV$^4$. The composition of the Universe thus obtained is consistent with the observational data.
The problem of perturbative reheating and its effects on the evolution of the curvature perturbations in tachyonic inflationary models was investigated in \cite{Re10}. It was found that reheating does not affect the amplitude of the curvature perturbations, and, after the transition, the relative non-adiabatic pressure perturbation decay extremely rapidly during the early stages of the radiation dominated epoch. These behavior ensure that the amplitude of the curvature perturbations remain unaffected during reheating.
It was pointed out in \cite{Re9} that among primordial magnetogenesis models, inflation is a prime candidate to explain the current existence of cosmological magnetic fields. Their energy density decreases as radiation during the decelerating eras of the universe, and in particular during reheating. Avoiding magnetic field backreaction is always complementary to CMB, and can give stronger limits on reheating for all high energy models of inflation.
The reheating dynamics after the inflation induced by $R^2$-corrected $f(R)$ models were considered \cite{Re8}. The inflationary and reheating dynamics was analyzed in the Einstein frame. Observational constraints on the model were also discussed.

The processes of particle production from the inflaton, their subsequent thermalization and evolution of inflaton/plasma system by taking dissipation of the inflaton in a hot plasma into account were investigated in \cite{Re7}, and it was shown  that the reheating temperature is significantly affected by these effects.

In $f(R)$ modified gravity models  the reheating dynamics after the inflation is significantly modified,  and affects the shape of the gravitational wave background spectrum \cite{Re6}.
In \cite{Re5} it was shown that reheating considerations may provide additional constraints to some classes of inflationary  models.
The current Planck satellite measurements of the Cosmic Microwave Background  anisotropies constrain the kinematic properties of the reheating era for most of the inflationary models \cite{Re4}. This result is obtained by deriving the marginalized posterior distributions of the reheating parameter for about 200 models. It turns out that the precision of the current CMB data is such that estimating the observational performance of an inflationary model now requires to incorporate information about its reheating history.
It was pointed out in \cite{Re3} that the inflaton coupling to the Standard Model particles is generated radiatively, even if absent at tree level. Hence, the dynamics of the Higgs field during inflation can change dramatically.
The inflaton decay and reheating period after the end of inflation in the non-minimal derivative coupling gravity model with chaotic potential was investigated in \cite{Re2}. This model provides an enhanced slow-roll inflation, caused by gravitationally enhanced friction.
A scale-invariant model of quadratic gravity with a non-minimally coupled scalar field was studied in \cite{Re1}. At the end of inflation, the Hubble parameter and the scalar field converge to a stable fixed point through damped oscillations that can reheat the Universe in various ways. For reviews on the inflationary reheating see \cite{allahverdi2010reheating} and \cite{RehRev}, respectively.

If the importance of the nonequilibrium and dissipative aspects of particle creation during reheating has been already pointed out a long time ago \cite{Zim}, its irreversible and open character has not received a similar attention. Thermodynamical systems in which matter creation occurs belong to the class of open thermodynamical systems, in which the usual adiabatic conservation laws are modified by explicitly including irreversible matter creation \cite{prigogine1988thermodynamics}. The thermodynamics of open systems has been also applied first to cosmology in \cite{prigogine1988thermodynamics}. The explicit inclusion of the matter creation in the matter – energy stress tensor in the Einstein field equations leads to a three stage cosmological model, beginning
with an instability of the vacuum. During the first stage the Universe is driven from an initial fluctuation of the vacuum to a de Sitter space, and particle creation occur. The de Sitter space exists during the decay time of its constituents (second stage), and ends, after a phase transition, in the usual FRW phase. The phenomenological approach of \cite{prigogine1988thermodynamics}, was further discussed and generalised in \cite{Calv, Calv1} through a covariant formulation, allowing specific entropy variations as usually expected for nonequilibrium processes. The cosmological implications of the irreversible matter creation processes in the framework of the thermodynamics of open or quantum systems  have been intensively investigated in \cite{op1,op2,op3,op4,op5,op5a,op5b,op5c,op6,op7,op7a,op8,op9,op10,op10a, op11,op12,op13,op14,op15,op16, op17,op18, op19}.

From the analysis of some classes of scalar field potentials, including the $\phi ^1$ potential, which provide a good fit to the CMB measurements, in \cite{Reh1} it was found that the Planck 2015 68\% confidence limit upper bound on the spectral index, $n_s$ implies an upper bound on the reheating temperature of $T_{reh}\leq 6\times 10^{10}$ GeV. The observations also exclude instantaneous reheating. Hence the low reheating temperatures allowed by some scalar field models open the possibility that dark matter could have been produced during the reheating period, instead of the radiation dominated era in the history of the Universe. Such a possibility could lead to a major change in our understanding of the early evolution stages of the Universe, leading to different predictions for the relic density and momentum distribution of WIMPs, sterile neutrinos, and axions.

It is the purpose of the present paper to apply the thermodynamics of open
systems, as introduced in \cite{prigogine1988thermodynamics}  and \cite{Calv}, to a cosmological fluid mixture, in which particle decay and production
occur, consisting of two basic components: scalar field, and matter, respectively. From a cosmological point of view this physical situation is specific to the preheating/reheating period of the inflationary cosmological models. The thermodynamics of irreversible processes as applied to inflationary cosmological models leads to a self-consistent description of the matter creation processes, which determines the whole dynamics and future evolution of
the Universe. By combining the basic principles of the thermodynamics of open systems with the cosmological Einstein equations for a flat, homogeneous and isotropic Universe we obtain a set of first order ordinary differential equations describing matter creation due to the decay of the scalar field. As for the matter components we specifically consider a model in which ordinary matter is a mixture of two components, a relativistic one (radiation), and a pressureless fluid, assumed to correspond to dark matter. In order to simplify the analysis of the reheating equations we reformulate them in a dimensionless form, by introducing a set of appropriate dimensionless variables. As a cosmological application of the general formalism we investigate matter creation from a scalar field by adopting different mathematical forms of the self-interaction potential $V$ of the field. More exactly, we will consider power law, exponential and Higgs type potentials. For each case the evolution of the post-inflationary Universe with matter creation is investigated in detail by solving numerically the set of the cosmological reheating evolution equations.  The results display the process of inflationary scalar field decaying to matter and dark matter, and the effects of the expansion of the Universe on the decay. We concentrate on the evolution of the dark matter and radiation components, which increase from zero to a maximum value, as well as on the scalar field decay. A simple model, which can be fully  solved analytically, is also presented. As a general result we find that the scalar field potential also plays an important role in the reheating process, and in the decay of the scalar field. Some of the cosmological parameters of each model are also presented, and analyzed in detail.  The models are constrained by the observational parameters called the inflationary parameters, including the scalar spectral index $n_s$, the tensor to scalar ratio $r$, the number of e-folds $N$, and the reheating temperature $T_{\text{reh}}$. These parameters can be calculated  directly from the potentials used in the models, and they can be used together with the observational data to constrain the free parameters in the potentials.  The study of the reheating models, and the development if new formalisms and constraints to the theories are undoubtedly an essential part of the completion to the theory. Analysis of the current experimental data, including the calculation of the inflationary parameters, is able to test the reheating models.

The present paper is organized as follows. The basic results in the thermodynamics of open systems are briefly reviewed in Section~\ref{sect1}. The full set of equations describing matter creation due to the decay of a scalar field are obtained in Section~\ref{sect2}, where the observational constraints on the inflationary and reheating models are also presented. An exact solution of the system of the reheating equations, corresponding to simple form of the scalar field (coherent wave) is obtained in Section~\ref{sect3}. Several reheating models, corresponding to different choices of the scalar field self-interaction potential are considered in Section~\ref{sect4}, by numerically solving the set of reheating evolution equations. The time dynamics of the cosmological scalar field, and of the matter components are obtained, and analyzed in detail. We conclude and discuss our results in Section~\ref{sect5}.

\section{Irreversible thermodynamics of matter creation}\label{sect1}

In this Section, we briefly review the thermodynamics of matter creation in open systems, which we will use for the study of the post-inflationary reheating processes.  In our presentation we start from the fundamental laws of thermodynamics, we proceed to the general covariant formulation of the theory, and then we apply our results to the particular case of homogeneous and isotropic cosmological models.

\subsection{The First Law of Thermodynamics}

The main novel element that appears in the thermodynamics of open systems with energy/matter exchange is the creation pressure, related to the variation of the particle number. The expression of the creation pressure can be derived from the fundamental laws of thermodynamics \cite{prigogine1988thermodynamics}. We consider a thermodynamic system consisting of $\mathcal{N}$ particles,  inside a volume $\mathcal{V}$. For a closed system, $\mathcal{N}$ is a constant, and the first law of thermodynamics - the energy conservation equation - is
\begin{equation}
{\rm d}\mathcal{E}={\rm d}\mathcal{Q}-p{\rm d}\mathcal{V},
\end{equation}
where $\mathcal{E}$ is the internal energy, $\mathcal{Q}$ is the heat received by the system, and $p$ is the thermodynamic pressure. We can also write the energy conservation as
\begin{equation}
{\rm d}\left( \frac{\rho}{n}\right) ={\rm d}q-p{\rm d}\left( \frac{1}{n}\right),
\end{equation}
where the energy density $\rho=\mathcal{E}/\mathcal{V}$, the particle number density $n=\mathcal{N}/\mathcal{V}$, and ${\rm d}q={\rm d}\mathcal{Q}/\mathcal{N}$, respectively. Adiabatic transformations $({\rm d}\mathcal{Q}=0)$ are described by
\begin{equation}\label{closefirst}
{\rm d}(\rho \mathcal{V})+p{\rm d}\mathcal{V}=0.
\end{equation}

For an open system, the particle number can vary in time, so that $\mathcal{N}=\mathcal{N}(t)$. By including the thermodynamic parameters of the newly created particles, the energy conservation equation is modified to \cite{prigogine1988thermodynamics}
\begin{equation}
{\rm d}(\rho \mathcal{V})={\rm d}\mathcal{Q}-p{\rm d}\mathcal{V}+\frac{h}{n}{\rm d}(n\mathcal{V}),
\end{equation}
where $h=\rho+p$ is the enthalpy per unit volume. Adiabatic transformations are then expressed as
\begin{equation}\label{openfirst}
{\rm d}(\rho \mathcal{V})+p{\rm d}\mathcal{V}-\frac{h}{n}{\rm d}(n\mathcal{V})=0.
\end{equation}
The matter creation process is a source of the internal energy of the system. From Eq.~(\ref{openfirst}) we obtain the relation
\begin{equation}\label{relarhon}
\dot{\rho}=(\rho+p)\frac{\dot{n}}{n},
\end{equation}
describing the variation of the matter density due to the creation processes. On the other hand the above equation can also describe physical systems in the absence of matter creation, like, for example, the photon gas,  in which, for a particular temperature, the particle number $N$ varies with the volume in a fixed manner, adjusting itself to have a constant density of photons. Hence, for a photon gas the particle number $N$ is a variable, and not a constant quantity, like in an ordinary gas.  The physical mechanism by which thermodynamic equilibrium can be established consists in the absorbtion and emission of photons by matter. For a radiation like equation of state, with $p=\rho /3$, from Eq.~(\ref{relarhon}) we obtain $\rho =\rho _0\left(n/n_0\right)^{4/3}$, with $\rho _0$ and $n_0$ constants,  giving the well-known energy density - particle number relation for the ultra-relativistic photon gas.

Alternatively, we can reformulate the law of the energy conservation by introducing the matter creation pressure, so that the energy conservation equation for open systems takes a form similar to Eq.~(\ref{closefirst}),  with the pressure written as $\tilde{p}=p+p_c$,
\begin{equation}
{\rm d}(\rho\mathcal{V})=-(p+p_c){\rm d}\mathcal{V},
\end{equation}
where the creation pressure associated describing the creation of the matter is
\begin{equation}
p_c=-\frac{\rho+p}{n}\frac{{\rm d}(n\mathcal{V})}{{\rm d}\mathcal{V}}.
\end{equation}

\subsection{The Second Law of Thermodynamics}

 In order to formulate the second law of thermodynamics for open systems, we define first the entropy flow ${\rm d_e}\mathcal{S}$, and the entropy creation ${\rm d_i}\mathcal{S}\geq 0$, so that the total entropy variation of the system is written as \cite{prigogine1988thermodynamics}
\begin{equation}
{\rm d}\mathcal{S}={\rm d_e}\mathcal{S}+{\rm d_i}\mathcal{S}\geq 0.
\end{equation}
For a closed and adiabatic system, ${\rm d_i}\mathcal{S}\equiv 0$. For an open system with matter creation, in the following we assume that  matter is created in a thermal equilibrium state ${\rm d_e}\mathcal{S}=0$. Therefore the entropy increases only due to the particle creation processes.

The total differential of the entropy is
\bea
T{\rm d}(s\mathcal{V})&=&{\rm d}(\rho \mathcal{V})-\frac{h}{n}{\rm d}(n\mathcal{V})+T{\rm d}(s\mathcal{V})=\nonumber\\
&&{\rm d}(\rho \mathcal{V})-(h-Ts){\rm d}\mathcal{V}
={\rm d}(\rho \mathcal{V})-\mu{\rm d}(n\mathcal{V}),\nonumber\\
\eea
where the entropy density $s=\mathcal{S}/\mathcal{V}\geq 0$, and the chemical potential $\mu=(h-\mathcal{T}s)/n\geq 0$. Hence the entropy production for an open system is
\begin{equation}\label{entropypro}
T{\rm d_i}\mathcal{S}=T{\rm d}\mathcal{S}=\left( \frac{h}{n}\right) {\rm d}(n\mathcal{V})-\mu{\rm d}(n\mathcal{V})
=T\left( \frac{s}{n}\right) {\rm d}(n\mathcal{V})\geq 0
\end{equation}

The matter creation processes can also be formulated in the framework of general relativity. In order to apply the thermodynamics of open systems to cosmology, in the following we consider the formulation of the thermodynamics of open systems  in a general relativistic covariant form \cite {Calv}. Then, by starting from this formulation, we  particularize it to the case of a homogeneous and isotropic Universe, by also deriving the entropy of the system \cite{op16}.

\subsection{General relativistic description of the thermodynamic of open systems}

In the following we denote the energy-momentum tensor of the system as $\mathcal{T}^{\mu\nu}$, and we introduce the entropy flux vector $s^{\mu}$, the particle flux vector $\mathcal{N}^{\mu}$, and the four-velocity $u^{\mu}$ of the relativistic fluid. These variables are defined as
\begin{equation}\label{enmotens}
\mathcal{T}^{\mu\nu}=(\rho+p+p_c)u^{\mu}u^{\nu}-(p+p_c)g^{\mu\nu},
\end{equation}
\begin{equation}\label{entropyflux}
s^{\mu}=n\sigma u^{\mu},
\mathcal{N}^{\mu}=nu^{\mu},
\end{equation}
where $p_c$ is the matter creation pressure, $\sigma$ is the specific entropy per particle, and $n$ is the particle number density.

The energy conservation law requires
\begin{equation}\label{econs}
\nabla_{\nu}\mathcal{T}^{\mu\nu}=0,
\end{equation}
while the second law of thermodynamics imposes the constraint
\begin{equation}
\nabla_{\mu}s^{\mu}\geq 0,
\end{equation}
on the entropy flux four-vector. The balance equation of the particle flux vector is given by
\begin{equation}\label{baparflux}
\nabla_{\mu}\mathcal{N}^{\mu}=\Psi,
\end{equation}
 where $\Psi>0$ represents a particle source, or a particle sink if $\Psi<0$. For an open system, the Gibbs equation is \cite{Calv}
\begin{equation}\label{gibbseq}
nT{\rm d}\sigma={\rm d}\rho-\frac{\rho+p}{n}{\rm d}n.
\end{equation}
In order to derive the energy balance equation  we multiply both sides of Eq.~(\ref{econs}) by $u^{\mu}$, thus obtaining
\bea
u_{\mu}\nabla_{\nu}\mathcal{T}^{\mu\nu}&=&u_{\mu}\nabla_{\nu}\left(\rho+p+p_c\right)u^{\mu}u^{\nu}
+u_{\mu}\left(\rho+p+p_c\right)\times \nonumber\\
&&\nabla_{\nu}(u^{\mu}u^{\nu})
-u_{\mu}\nabla^{\mu}\left(p+p_c\right)\nonumber\\
&=&u^{\nu}\nabla_{\nu}(\rho+p+p_c)
+\left(\rho+p+p_c\right)\times\nonumber\\
&&(u_{\mu}u^{\nu}\nabla_{\nu}u^{\mu}+u_{\mu}u^{\mu}\nabla_{\nu}u^{\nu})
-\left(\dot{p}+\dot{p_c}\right)\nonumber\\
&=&\dot{\rho}+\left(\rho+p+p_c\right)\nabla_{\nu}u^{\nu}=0,
\eea
where we have used the relations $\dot{\rho}=u^{\mu}\nabla_{\mu}\rho={\rm d}\rho/{\rm d}s$, $u_{\mu}u^{\mu}=1$, and $u_{\mu}u^{\nu}\nabla_{\nu}u^{\mu}=0$, respectively. Hence we have obtained the energy conservation equation for open systems as
\begin{equation}\label{dotrho}
\dot{\rho}+(\rho+p+p_c)\nabla_{\mu}u^{\mu}=0.
\end{equation}

In order to obtain the entropy variation we substitute the relation (\ref{dotrho}) into the Gibbs equation (\ref{gibbseq}), and we use Eqs.~(\ref{entropyflux}) and (\ref{baparflux}) to obtain first
\bea
nTu^{\mu}\nabla_{\mu}\sigma&=&\dot{\rho}-\frac{\rho+p}{n}u^{\mu}\nabla_{\mu}n
T\nabla_{\mu}s^{\mu}-T\sigma\nabla_{\mu}\mathcal{N}^{\mu} \nonumber\\
&=&-(\rho+p+p_c)\nabla_{\mu}u^{\mu}-\frac{\rho+p}{n}\times \nonumber\\
&&\left(\nabla_{\mu}\mathcal{N}^{\mu}-n\nabla_{\mu}u^{\mu}\right)
T\nabla_{\mu}s^{\mu}\nonumber\\
&=&-p_c\nabla_{\mu}u^{\mu}-\left( \frac{\rho+p}{n}-T\sigma\right) \nabla_{\mu}\mathcal{N}^{\mu}.
\eea
Thus we obtain the entropy balance equation as
\begin{equation}\label{baentro}
\nabla_{\mu}s^{\mu}=-\frac{p_c\Theta}{T}-\frac{\mu\Psi}{T},
\end{equation}
where $\mu=(\rho+p)/n-T\sigma$ is the chemical potential, and $\Theta=\nabla_{\mu}u^{\mu}$ is the expansion of the fluid. From Eq.~(\ref{entropyflux}) we obtain
\begin{equation}\label{entroba}
\nabla_{\mu}s^{\mu}=\sigma u^{\mu}\nabla_{\mu}n+nu^{\mu}\nabla_{\mu}\sigma
=\sigma\Psi+n\dot{\sigma},
\end{equation}
where we have constrained the specific entropy per particle to be a constant, $\dot{\sigma}=0$. From Eqs.~(\ref{baentro}) and (\ref{entroba}) it follows that
\begin{equation}
-\frac{p_c\Theta}{T}-\frac{\mu\Psi}{T}=\sigma\Psi,
\end{equation}
and thus we obtain matter creation pressure as
\begin{equation}\label{pc}
p_c=-\frac{\rho+p}{n}\frac{\Psi}{\Theta}.
\end{equation}
One can see from the above expression that if there is no matter creation, $\Psi=0$, and then the matter creation pressure $p_c$ vanishes, and the entropy becomes a constant. If matter is created with $\Psi>0$ and $\Theta>0$, then the matter creation pressure will be negative, $p_c<0$.

\subsection{Matter creation in homogeneous and isotropic Universes}

In the case of a homogeneous and isotropic flat Friedmann-Robertson-Walker (FRW) geometry, the line element is given by
\begin{equation}\label{metr}
{\rm d}s^2={\rm d}t^2-a^2(t)\left({\rm d}x^2+{\rm d}y^2+{\rm d}z^2\right),
\end{equation}
where $a(t)$ is the scale factor. In the comoving frame, the four-velocity is given by $u^{\mu}=(1,0,0,0)$. For any thermodynamic function  $\mathcal{F}$ its derivative is given by $\dot{\mathcal{F}}=u^{\mu}\nabla_{\mu}\mathcal{F}={\rm d}\mathcal{F}/{\rm d}t$.
For the  covariant derivative of the four-velocity,  by taking into account that $\mathcal{V}=a^3$, we find
\bea
\nabla_{\mu}u^{\mu}=\frac{1}{\sqrt{-g}}\frac{\partial(\sqrt{-g}u^{\mu})}{\partial x^{\mu}}=\frac{1}{a^3}\frac{\partial(a^3u^{\mu})}{\partial x^{\mu}}=\frac{\dot{\mathcal{V}}}{\mathcal{V}}=3H,
\eea
where $H$, the  expansion of the fluid (the Hubble function) is given by $\nabla_{\mu}u^{\mu}=\dot{\mathcal{V}}/\mathcal{V}=\dot{a}/a$. The components of the energy-momentum tensor (\ref{enmotens}) in the presence of particle creation are
\begin{equation}
\mathcal{T}^0_0=\rho ,
\mathcal{T}^1_1=\mathcal{T}^2_2=\mathcal{T}^3_3=-(p+p_c).
\end{equation}

From Eq.~(\ref{baparflux}) we obtain the particle number balance equation as
\begin{equation}
\Psi=u^{\mu}\nabla_{\mu}n+n\nabla_{\mu}u^{\mu}=\dot{n}+n\frac{\dot{\mathcal{V}}}{\mathcal{V}}.
\end{equation}
Hence, by substituting  the above equation into Eq.~(\ref{entropypro}), we obtain the variation of the entropy of the open system due to particle creation processes as
\begin{equation}
\mathcal{S}=\mathcal{S}_0\exp{\left( \int_{t_0}^t \frac{\Psi(t)}{n}{\rm d}t\right) },
\end{equation}
where $\mathcal{S}_0=\mathcal{S}\left(t_0\right)$ is an arbitrary integration constant.

\section{Reheating Dynamics: irreversible open systems thermodynamics description}\label{sect2}

Reheating is the period at the end of inflation when matter particles were created through the decay of the inflationary scalar field. The particle creation is essentially determined  by  the energy density $\rho _{\phi}$ and pressure $p_{\phi}$ of the inflationary scalar field $\phi $. For all inflationary models, inflation begins at the energy scale of order $10^{14}$ GeV, or, equivalently,  at an age of the Universe of the order of $10^{-34}$ s \cite{B31}. Hence reheating starts at the end of inflation, at a time of the order of $10^{-32}$ s, and it should end before the hot big bang at $10^{-18}$ s, which means that the energy scale of reheating is from around $10^{-7}$ GeV up to $10^7$ GeV. The reheating temperature is an important parameter of the model, which we will discuss in detail.

In this Section we derive first the general equations describing the reheating process in the framework of the thermodynamics of open systems. We also review the main observational constraints and parameters of the inflationary cosmological models, prior and during the reheating.

\subsection{Irreversible thermodynamical description of reheating}

We consider the Universe in the latest stages of inflation as an open thermodynamic system with matter creation. The three constituents of the Universe are the inflationary scalar field, the ordinary matter, with energy density $\rho_{\text{m}}$ and pressure $p_{\text{m}}$, and the dark matter, with energy density $\rho_{\text{DM}}$ and pressure $p_{\text{DM}}$, respectively. Both the ordinary matter and the dark matter are assumed to be perfect fluids. The total energy density $\rho$ of the system is
\begin{equation}
\rho=\rho_{\phi}+\rho_{\text{DM}}+\rho_{\text{m}}.
\end{equation}
Similarly, the total pressure $p$ and the total particle number density $n$ of the Universe at the end of inflation  are given by
\begin{equation}
p=p_{\phi}+p_{\text{DM}}+p_{\text{m}},
n=n_{\phi}+n_{\text{DM}}+n_{\text{m}}.
\end{equation}

In the presence of particle creation processes, the stress-energy tensor $\mathcal{T}^{\mu\nu}$ of the inflationary Universe in obtained as
\begin{equation}
\mathcal{T}^{\mu\nu}=\left(\rho +p+p_c^{(\rm total)}\right)u^{\mu}u^{\nu}-(p+p_c^{(\rm total)})g^{\mu\nu},
\end{equation}
where $p_c^{(\rm total)}$ is the total creation pressure describing matter creation, and we have assumed that all three fluid components are comoving, thus having the same four velocity $u^{\mu}$. In the following we will restrict our analysis to the case of the flat  Friedmann-Robertson-Walker Universe, with line element given by Eq.~(\ref{metr}). The basic equations describing the reheating process in the framework of the thermodynamic of open systems can be formulated as follows.

\paragraph{The particle number balance equation}

In the FRW geometry the balance equation of the particle flux vector, Eq.~(\ref{baparflux}), can be written for each species of particle with particle numbers $n_{\phi}$, $n_{{\rm m}}$, $n_{{\rm DM}}$ as
\begin{equation}
\Psi _a=\dot{n}_a+n_a\nabla_{\mu}u^{\mu}=\dot{n}_a+3Hn_a, a={\phi},{\rm m}, {\rm DM}.
\end{equation}

We assume that {\it the particle source function $\Psi$  is proportional to the energy density of the inflationary scalar field $\rho_{\phi}$},  with the proportionality factor $\Gamma $ representing a decay (creation) width. Hence the balance equations of the particle number densities are given by
\begin{equation}
\dot{n}_{\phi}+3Hn_{\phi}=-\frac{\Gamma_{\phi}}{m_{\phi}}\rho_{\phi},
\end{equation}
\begin{equation}
\dot{n}_{\text{DM}}+3Hn_{\text{DM}}=\frac{\Gamma_{\text{DM}}}{m_{\rm DM}}\rho_{\phi},
\end{equation}
\begin{equation}
\dot{n}_{\text{m}}+3Hn_{\text{m}}=\frac{\Gamma_{\text{m}}}{m_{\rm m}}\rho_{\phi},
\end{equation}
where $\Gamma_{\phi}$, $\Gamma_{\text{DM}}$ and $\Gamma_{\text{m}}$ are generally different, nonzero functions, and $m_{\phi}$, $m_{{\rm DM}}$ and $m_{{\rm m}}$ denotes the masses of the scalar field, dark matter, and ordinary matter particles, respectively. For the whole system, the particle balance equation is
\begin{equation}
\dot{n}+3Hn=(\Gamma_{\text{m}}+\Gamma_{\text{DM}}-\Gamma_{\phi})\rho_{\phi}.
\end{equation}

\paragraph{The energy balance equations}

From Eq.~(\ref{relarhon}) we obtain the equations of the energy density balance of the cosmological fluids as
\begin{equation}\label{rhophieq}
\dot{\rho}_{\phi}+3H(\rho_{\phi}+p_{\phi})=-\frac{\Gamma_{\phi}(\rho_{\phi}+p_{\phi})\rho_{\phi}}{m_{\phi}n_{\phi}},
\end{equation}
\begin{equation}
\dot{\rho}_{\text{DM}}+3H(\rho_{\text{DM}}+p_{\text{DM}})=\frac{\Gamma_{\text{DM}}(\rho_{\text{DM}}+p_{\text{DM}})\rho_{\phi}}{m_{{\rm DM}}n_{\text{DM}}},
\end{equation}
\begin{equation}
\dot{\rho}_{\text{m}}+3H(\rho_{\text{m}}+p_{\text{m}})=\frac{\Gamma_{\text{m}}(\rho_{\text{m}}+p_{\text{m}})\rho_{\phi}}{m_{{\rm m}}n_{\text{m}}},
\end{equation}
and
\bea
\dot{\rho}+3H\rho&=&\rho_{\phi}\Bigg[ \frac{\Gamma_{\text{m}}(\rho_{\text{m}}+p_{\text{m}})}{m_{{\rm m}}n_{\text{m}}}+\frac{\Gamma_{\text{DM}}(\rho_{\text{DM}}+p_{\text{DM}})}{m_{{\rm DM}}n_{\text{DM}}}-\nonumber\\
&&\frac{\Gamma_{\phi}(\rho_{\phi}+p_{\phi})}{m_{\phi}n_{\phi}}\Bigg],
\eea
respectively.

\paragraph{The matter creation pressures}
From the expression of the creation pressure associated to the newly created particles Eq.~(\ref{pc}) we find
\begin{equation}
p_c^{(a)}=-\frac{\rho^{(a)}+p^{(a)}}{n^{(a)}}\frac{\Psi^{(a)}}{\Theta^{(a)}}=-\frac{(\rho^{(a)}+p^{(a)})\Psi^{(a)}}{3Hn^{(a)}},
\end{equation}
where subscript $(a)$ denotes the different cosmological fluids. During the reheating phase, for each component, the decay and creation pressures are
\begin{equation}
p_c^{(\phi)}=\frac{\Gamma_{\phi}(\rho_{\phi}+p_{\phi})\rho_{\phi}}{3Hm_{\phi}n_{\phi}},
\end{equation}
\begin{equation}
p_c^{(\text{DM})}=-\frac{\Gamma_{\text{DM}}(\rho_{\text{DM}}+p_{\text{DM}})\rho_{\phi}}{3Hm_{{\rm DM}}n_{\text{DM}}},
\end{equation}
\begin{equation}
p_c^{(\text{m})}=-\frac{\Gamma_{\text{m}}(\rho_{\text{m}}+p_{\text{m}})\rho_{\phi}}{3Hm_{{\rm m}}n_{\text{m}}}.
\end{equation}
The total creation pressure is
\bea
p_c^{(\rm total)}&=&\frac{\rho_{\phi}}{3H} \Bigg[ \frac{\Gamma_{\phi}}{m_{\phi}n_{\phi}}(\rho_{\phi}+p_{\phi})-\frac{\Gamma_ {{\rm DM}}}{m_{{\rm DM}}n_{\text{DM}}}\times \nonumber\\
&&\left(\rho_{\text{DM}}+p_{\text{DM}}\right)-
\frac{\Gamma_{\text{m}}}{m_{{\rm m}}n_{\text{m}}}(\rho_{\text{m}}+p_{\text{m}}) \Bigg].
\eea

\paragraph{The gravitational field equations}
The Einstein gravitational field equations in a flat FRW universe with inflationary scalar field, ordinary matter and dark matter fluids in the presence of matter creation can be written as follows
\begin{equation}\label{gather}
3H^2=\frac{1}{m_{Pl}^2}\left(\rho_{\phi}+\rho_{\text{DM}}+\rho_{\text{m}}\right),
\ee
\be
2\dot{H}+3H^2=-\frac{1}{m_{Pl}^2}\left(p_{\phi}+p_{\text{DM}}+p_{\text{m}}+p_c^{(\rm total)}\right).
\end{equation}
In the present paper we use natural units with $c =\hbar=k_B=1$, in which $8\pi G = 1/m_{Pl}^2 = 1.687 \times 10^{-43}$ MeV$^{-2}$, where $m_{Pl}$ is the ”reduced” Planck mass.
\paragraph{The entropy generation}
The entropy generated by the newly created matter is
\begin{equation}
\frac{\mathcal{S}_{\text{m}}(t)}{\mathcal{S}_{\text{m}}(t_0)}=\exp{\left( \int^t_{t_0}\Gamma_{\text{m}}\frac{\rho_{\phi}}{m_{{\rm m}}n_{\text{m}}}dt\right) }.
\end{equation}

\subsection{The models of reheating}

Presently, the accepted paradigm for
reheating is based on particle production and parametric amplification, as extensively
discussed  in \cite{allahverdi2010reheating} and \cite{RehRev}. According to this model,  the main process of particle production is parametric
amplification in combination with a slowing expansion. This mechanism
entails a description in terms of Mathieu equations, either for the
scalar or fermion fields. These equations feature unstable bands with
the concomitant parametric instabilities. This mechanism cannot be
described in terms of a simple $\Gamma$ functions, and different bands feature
different Floquet exponents (in Minkowski or near Minkowski space
time), and correspond to different wave-vector bands.  In this picture, matter is
created in a highly non-thermal state (determined primarily by the
position and widths of the unstable parametric bands), and thermalizes
(at least in principle, although never quite proven directly) via scattering.
This last step requires a clear knowledge of relaxation times, which
entail understanding the microscopic processes and cross sections.

In the following we will consider models of reheating involving energy transfer between the inflationary scalar field, and the newly created ordinary matter, and dark matter, respectively. We assume that particles are created through an irreversible thermodynamic process. In this Subsection, we present explicitly  the general equations governing the matter and dark matter creation during inflationary reheating.
In our reheating models,  the energy density and pressure for the inflationary scalar field have the form
\begin{equation}\label{rhophi}
\rho_{\phi}=\frac{1}{2}\dot{\phi}^2+V(\phi),
p_{\phi}=\frac{1}{2}\dot{\phi}^2-V(\phi),
\end{equation}
where $V(\phi)$ is the self-interaction potential of the scalar field. Substituting Eqs.~(\ref{rhophi}) into Eq.~(\ref{rhophieq}), we obtain the generalized Klein-Gordon equation for the scalar field in the presence of matter creation. The equation of state of the dark matter is defined as $
w_{\text{DM}}=p_{\text{DM}}/\rho_{\text{DM}}\equiv 0$, since we assume that the pressure of the newly created dark matter particles is $p_{\text{DM}}=0$.
The energy density of dark matter is proportional to its particle number density, $\rho_{\text{DM}}\sim n_{\text{DM}}$, so that
$\rho_{\text{DM}}=m_{\text{DM}}n_{\text{DM}}$, where $m_{\text{DM}}$ is the mass of the dark matter particles. For pressureless matter from Eq.~(\ref{relarhon}) we obtain the general relation
\begin{equation}
n\dot{\rho}={\rho}\dot{n}.
\end{equation}

The newly created matter is considered to be in form of radiation (an ultra-relativistic fluid), with the equation of state
$w_{\text{m}}=p_{\text{m}}/\rho_{\text{m}}=1/3$. The relation between the energy density and the particle number density of radiation is $\rho_{\text{m}}\sim n_{\text{m}}^{4/3}$.

In summary, the set of equations governing the matter creation process during reheating is written below as
\be\label{geneqset}
3H^2= \frac{1}{m_{Pl}^2}\left(\rho_{\phi}+\rho_{\text{DM}}+\rho_{\text{m}}\right),
\ee
\be
2\dot{H}+3H^2=-\frac{1}{m_{Pl}^2}\left(p_{\phi}+p_{\text{m}}+p_c^{\text{(total)}}\right),
\ee
\be\label{KG}
\ddot{\phi}+V'(\phi)+3H\dot{\phi}+\frac{\Gamma_{\phi}\rho_{\phi}}{m_{\phi}n_{\phi}}\dot{\phi}=0,
\ee
\be
\dot{\rho}_{\text{DM}}+3H\rho_{\text{DM}}=\Gamma_{\text{DM}}\rho_{\phi},\quad \dot{\rho}_{\text{m}}+4H\rho_{\text{m}}=\Gamma_{\text{m}}\rho_{\phi},
\ee
\be\label{geneqsetf}
\dot{n}_{\phi}+3Hn_{\phi}+\frac{\Gamma_{\phi}}{m_{\phi}}\rho_{\phi}=0,
\ee
where $\Gamma_{\phi}$, $\Gamma_{\text{DM}}$, $\Gamma_{\text{m}}$  are considered as constants.  Eq.~(\ref{KG}) is the generalized Klein-Gordon equation, describing the evolution of the scalar field.
The last term on the left hand side of this equation represents an effective friction term,  describing the effects of irreversible matter creation (decay) on the scalar field dynamics. If $\rho _{\phi}/n_{\phi}={\rm constant}$, we recover Eq.~(\ref{phen}), introduced on a purely phenomenological basis in \cite{Alb}.

\subsubsection{Dimensionless form of the reheating equations}

In order to simplify the mathematical formalism we introduce a set of dimensionless variables $\tau$, $h$, $r_{\phi}$, $r_{\text{DM}}$, $r_{\text{m}}$, $N_{\phi}$, $N_{\text{m}}$, $\gamma _{\text{DM}}$, $\gamma _{\text{m}}$, and $v(\phi)$, respectively, defined as
\bea
&&t=\frac{1}{M_{\phi}}\tau,H=M_{\phi} h,\rho_a=M_{\phi} ^2m_{Pl}^2r_a,p_a=M_{\phi} ^2m_{Pl}^2P_a, \nonumber\\
&&a=\phi,\text{DM}, \text{m},c,
v(\phi)=\frac{1}{M_{\phi} ^2}V(\phi), n_a=M_{\phi} ^2N_a, \nonumber\\
&&a=\phi,\text{DM}, \text{m},
\Gamma _b/m_b=M_{\phi} \gamma _b,
b=\phi, \text{DM}, \text{m},
\eea
where $M_{\phi}$ is a constant with the physical dimension of a mass.

In the new variables the equations (\ref{geneqset})-(\ref{geneqsetf}) take the dimensionless form
\be\label{d1}
3h^2=r_{\phi}+r_{\text{DM}}+r_{\text{m}},
\ee
\be\label{d2}
2\frac{dh}{d\tau}+3h^2=-P_{\phi}-P_{\text{DM}}-P_c^{\text{(total)}},
\ee
\be\label{d3}
\frac{d^2\phi}{d\tau ^2}+v'(\phi)+3h\frac{d\phi}{d\tau}+\frac{\gamma _{\phi}r_{\phi}}{N_{\phi}}\frac{d\phi}{d\tau}=0,
\ee
\be\label{d4}
\frac{dr_{\text{DM}}}{d\tau}+3hr_{\text{DM}}=\gamma _{\text{DM}}r_{\phi},\frac{dr_{\text{m}}}{d\tau}+4hr_{\text{m}}=\gamma _{\text{m}}r_{\phi},
\ee
\be\label{d5}
\frac{dN_{\phi}}{d\tau}+3hN_{\phi}+\gamma _{\phi}r_{\phi}=0.
\ee

Eqs.~(\ref{d4}) and (\ref{d5}) can be solved exactly, and we obtain the following exact representation of the scalar field particle number and matter densities,
\be
N_{\phi}(\tau )=N_{\phi}(0)\frac{a(0)}{a^3(\tau )}-\frac{\int{\gamma _{\phi}\rho _{\phi}a^3(\tau)d\tau}}{a^3(\tau )},
\ee
\be
r_i=\frac{\int{\gamma _{a}\rho _{\phi}a^3(\tau)d\tau}}{a^n(\tau )},\left\{i={\rm DM}, n=3\right\}, \left\{i={\rm m}, n=4,\right\}.
\ee
When $\rho _{\phi}\rightarrow 0$, the particles creation processes end, and the Universe enters in the standard expansionary phase, with its chemical composition consisting of two matter components, as well as a possible background of scalar field particles. After reheating, the energy density of the dark matter and radiation varies as
\be
r_{{\rm DM}}=\frac{r_{{\rm DM0}}}{a^3}, r_{{\rm m}}=\frac{r_{{\rm m0}}}{a^4},
\ee
where $r_{{\rm DM0}}$ and $r_{{\rm m0}}$ are the dark matter and radiation energy densities at the end of reheating. The evolution of the scale factor is described by the equation
\be
h(a)=\frac{1}{a}\frac{da}{d\tau}=\frac{1}{\sqrt{3}}\sqrt{\frac{r_{{\rm DM0}}}{a^3}+\frac{r_{{\rm m0}}}{a^4}},
\ee
giving
\be
\tau (a)-\tau_0=\frac{2 \left(a r_{\text{DM0}}-2 r_{\text{m0}}\right) \sqrt{ r_{\text{DM0}}a+r_{\text{m0}}}}{\sqrt{3}
   r_{\text{DM0}}^2}.
\ee
The deceleration parameter for the post-reheating phase is given by
\be
q(a)=\frac{3  r_{\text{DM0}}a+4 r_{\text{m0}}}{2  \left(a r_{\text{DM0}}a+r_{\text{m0}}\right)}-1.
\ee
In the limit $ r_{\text{DM0}}a>>r_{\text{m0}}$, the deceleration parameter tends to the value $q=1/2$, indicating a decelerating expansion of the Universe.

\subsection{Observational tests of inflationary models}

In order to facilitate the comparison of the theoretical models of the reheating based on the thermodynamics of open systems  with the observational results, we also present and calculate various cosmological parameters, either directly from the inflationary models, or from the potentials of the inflationary scalar field.

\paragraph{Slow-roll parameters}

The slow-roll parameters were introduced in 1992, and are based on the slow-roll approximation \cite{liddle1992cobe}. Their calculations are presented in \cite{B1}. During the period before reheating, the inflationary scalar field has not yet decayed, and the field equation and the equation of the scalar field can be written as
\begin{equation}
3H^2=\frac{1}{m_{Pl}^2}\rho_{\phi}=\frac{1}{m_{Pl}^2}\left(\frac{1}{2}\dot{\phi}^2+V(\phi)\right),
\ee
\be
\ddot{\phi}+V'(\phi)+3H\dot{\phi}=0.
\end{equation}
The slow-roll approximation requires that the terms $\dot{\phi}^2/2$ and $\ddot{\phi}$ are sufficiently small, so that the above equations can be simplified to
\begin{equation}
3H^2\simeq \frac{1}{m_{Pl}^2}V(\phi),\quad
3H\dot{\phi}\simeq -\frac{1}{m_{Pl}^2}V'(\phi).
\end{equation}
The slow-roll parameters are defined as $\epsilon (\phi)=(1/2)\left(V'/V\right)^2$, and $\eta(\phi)=V''/V- (1/2)\left(V'/V\right)^2$, respectively \cite{B1,liddle1992cobe}.
The slow-roll approximation only holds when the following conditions are satisfied, $\epsilon(\phi)\ll 1$ and $\eta(\phi)|\ll 1$, respectively.

\paragraph{Inflationary Parameters}

During inflation, the Universe expands exponentially. The scale of inflation is represented by the scale factor $a$. The ratio of the scale factor at the end of inflation and at its beginning is always large, so that the logarithm of this quantity is of physical interest. The number of e-folds is defined as $N(t)=\ln\left[a\left(t_{\text{end}}\right)/a(t)\right]$, where $t_{\text{end}}$ is the time when inflation ends. The number of e-folds can also be obtained from the comoving Hubble length \cite{B1}. Under the slow-roll approximation, it can be calculated as
\begin{equation}
N=\int^{t_{\text{end}}}_{t}H{\rm d}t\simeq \int^{\phi_{\text{CMB}}}_{\phi_{\text{end}}}\frac{V}{V'}{\rm d}\phi=\int^{\phi_{\text{CMB}}}_{\phi_{\text{end}}}\frac{1}{\sqrt{\epsilon\left(\phi'\right)}}{\rm d}\phi',
\end{equation}
where $\phi_{\text{end}}$ is defined as $\epsilon(\phi_{\text{end}})=1$, when the slow-roll approximation is violated, and inflation ends. Typically $N\simeq 50$. Sometimes the integral is multiplied by a factor of $1/\sqrt{2}$.

The scalar spectral index $n_s$ is defined by the slow-roll parameters
$n_s=1-4\epsilon\left(\phi_{\text{CMB}}\right)+2\eta\left(\phi_{\text{CMB}}\right)$.
The index $n_s$ is usually a function of the parameters of the model. The value of $n_s$ can be obtained from the observations \cite{Pl4}, and is used to constrain the inflationary models. The tensor-to-scalar ratio $r$ is similarly defined as
$r=13.7\times \epsilon\left(\phi_{\text{CMB}}\right)$.
One can see from their definition that the two parameters $n_s$ and $r$ are not independent.

\paragraph{Cosmological parameters}

The deceleration parameter is defined as
\begin{equation}
q=\frac{{\rm d}}{{\rm d}t}\left( \frac{1}{H}\right) -1=-\frac{a\ddot{a}}{\dot{a}^2}.
\end{equation}
It indicates the  nature of the acceleration of the expansion of the Universe. $q<0$ means that the expansion of the Universe is accelerating, while a positive $q$ indicates deceleration.

The density parameters of the inflationary scalar field, matter and dark matter are defined as
$\Omega_{\phi}=\rho_{\phi}/\rho $, $\Omega_{\text{DM}}=\rho_{\text{DM}}/\rho $, $\Omega_{\text{m}}=\rho_{\text{m}}/\rho $.
The evolution of density parameters of each matter component in the early Universe gives us a hint on the matter creation process during inflationary reheating.

\subsubsection{Model parameters}

In studying the reheating models, we consider different scalar field potentials that influence the dynamics of the early Universe. During inflation the energy density of the scalar field is potential energy  dominated.

\paragraph{Scalar field potentials} We picked six potentials of cosmological interest we are going to investigate, and presented them  in Table~\ref{potential}.

\begin{widetext}
\begin{center}
\begin{table}[t]
\centering
\begin{tabular}{|c|c|c|}
\hline
Number & Potentials & Expressions \\
\hline
1 & Large-field Polynomial Potentials & $M_{\phi}^2\left( \frac{\phi}{\mu}\right) ^p,p>1$\\
\hline
2 & Small-field Polynomial Potentials & $M_{\phi}^2\left[ 1-\left( \frac{\phi}{\mu}\right) ^p\right] ,\phi\ll\mu,p>2$\\
\hline
3 & Small-field Quadratic Potentials & $M_{\phi}^2\left[ 1-\left( \frac{\phi}{\mu}\right) ^2\right] ,\phi\ll\mu$\\
\hline
4 & Linear Potentials & $M_{\phi}^2\left( \frac{\phi}{\mu}\right) $ and $M_{\phi}^2\left( 1-\frac{\phi}{\mu}\right) $\\
\hline
5 & Exponential Potentials & $M_{\phi}^2\exp{\left( \pm\sqrt{\frac{2}{p}}\phi^2\right) },p>0$\\
\hline
6 & Higgs Potentials & $M_{\phi}^2\left[ 1-\left( \frac{\phi}{\mu}\right) ^2+\left( \frac{\phi}{\nu}\right) ^4\right] $\\
\hline
\end{tabular}
\caption{Scalar field potentials}\label{potential}
\end{table}
\end{center}
\end{widetext}

Some of these potentials are simple, and satisfy the slow-roll conditions, like the large-field, small-field, and linear potentials. However, since the slow-roll condition is a sufficient  but not necessary condition for inflation \cite{B1}, some other forms of the potentials are also considered. We have also selected the well-studied exponential potential, which is easy to analyze and constrain observationally. Finally, as the only scalar field in the Standard Model of particle physics, the Higgs type potentials are also worth to investigate. The observational constraints of the six potentials are not exactly the same. Each potential has its own slow-roll and inflationary parameters, while they share similar constraints from the cosmological parameters and the reheating temperature. The values of the constants are discussed in the next paragraphs.

\paragraph{Slow-roll parameters}

The slow-roll parameters are obtained for each potential. Their values are shown in Table~\ref{slop}.

\begin{widetext}
\begin{center}
\begin{table}[t]
\centering
\begin{tabular}{|c|c|c|c|c|}
\hline
Number & $\phi_{\text{end}}^2$ & $\phi_{\text{CMB}}^2$ & $\epsilon(\phi_{\text{CMB}})$ & $\eta(\phi_{\text{CMB}})$\\
\hline
1 & $\frac{p^2}{2}$ & $\frac{p(p+200)}{2}$ & $\frac{p}{p+200}$ & $\frac{p-2}{p+200}$\\
\hline
2 & $\mu^2\left( \frac{\mu}{p}\right) ^{\frac{2}{p-1}}$ & $\mu^2\left[ \frac{\mu^2}{50p(p-2)}\right] ^{\frac{2}{p-2}}$ & $\simeq 0$ & $-\frac{p-1}{50(p-2)}$ \\
\hline
3 & / & $\phi_{\text{end}}^2\exp{(-\frac{200}{\mu^2})}$ & $\simeq 0$ & $-\frac{2}{\mu^2}$ \\
\hline
4 & / & / & $\frac{1}{2\mu^2}$ & $-\frac{1}{2\mu^2}$ \\
\hline
5 & / & / & $\frac{1}{p}$ & $\frac{1}{p}$ \\
\hline
6 & / & $\phi_{\text{end}}^2\exp{(-\frac{200}{\mu^2})}$ & $\simeq 0$ & $-\frac{2}{\mu^2}$ \\
\hline
\end{tabular}
\caption{Slow-roll parameters for each model~\cite{dodelson1997cosmic}}\label{slop}
\end{table}
\end{center}
\end{widetext}

\paragraph{Inflationary Parameters}
Similarly, the inflationary parameters, scalar spectral index $n_s$ and tensor-to-scalar ratio $r$ are presented in Table~\ref{infp}.

\begin{table}[t]
\centering
\begin{tabular}{|c|c|c|c|c|}
\hline
Number & \qquad & $n_s$ & \qquad  & $r$ \\
\hline
1 & \qquad  & $1-\frac{2p+4}{p+200}$ & \qquad  & $13.7\frac{p}{p+200}$\\
\hline
2 & \qquad  & $1-\frac{p-1}{25(p-2)}$ & \qquad  & $\simeq 0$ \\
\hline
3 & \qquad  & $1-\frac{4}{\mu^2}$ & \qquad  & $\simeq 0$\\
\hline
4 & \qquad  & $1-\frac{3}{\mu^2}$ & \qquad  & $\frac{13.7}{2\mu^2}$ \\
\hline
5 & \qquad  & $1-\frac{2}{p}$ & \qquad  & $\frac{13.7}{p}$\\
\hline
6 & \qquad  & $1-\frac{4}{\mu^2}$ & \qquad  & $\simeq 0$\\
\hline
\end{tabular}
\caption{Inflationary parameters for each model}\label{infp}
\end{table}

\paragraph{Mass scale $M_{\phi}$}

In order to compare the different potentials we assume that in all of them the mass scale $M_{\phi}$ is the same. We constrain  $M_{\phi}$ in the potentials by estimating the values of $V$ at the beginning and ending time of reheating. Reheating begins at the end of inflation at around $10^{-32}$ s and ends before $10^{-16}$ s \cite{B31}. Thus the value of the mass scale $M_{\phi}$ is
\bea
&&10^{18}{\rm \ s}^{-1}\leq M_{\phi} \leq 10^{32}{\rm \ s}^{-1},\nonumber\\
&&6.58\times 10^{-7}{\rm \ GeV}\leq M_{\phi} \leq 6.58\times 10^{7}{\rm \ GeV}.
\eea
In the set of reheating equations, $M_{\phi}$ is defined as
\begin{equation}
M_{\phi}=\frac{\tau}{t}=\frac{\tau}{t(\text{s})}\times 6.58\times 10^{-25}{\rm \ GeV}=0.658\tau{\rm \ keV}.
\end{equation}
When the dimensionless time variable is in the range $0.3\leq \tau \leq 3$, the value of $M_{\phi}$ varies as
\begin{equation}
0.1974{\rm \ keV}\leq M_{\phi} \leq 1.974{\rm \ keV}.
\end{equation}
In our irreversible thermodynamic study of reheating models we adopt the value $M_{\phi}\approx 1{\rm \ keV}$.

\paragraph{Reheating Temperature}\label{treh}

The reheating temperature is usually referred to as the maximum temperature during reheating period. However, it was argued that the reheating temperature is not necessarily the maximum temperature of the Universe \cite{scherrer1985decaying}. In the radiation dominated Universe the relation between energy density and the temperature of the matter (radiation) is $\rho_{\text{m}}\sim T^4$. With $\rho_{\text{m}}=M_{\phi} ^2m_{Pl}^2r_{\text{m}}$ we have
$T\sim M_{\phi}^{1/2}m_{Pl}^{1/2}\sqrt[4]{r_{\text{m}}}$.
When the dimensionless variable is approximately $0.1\leq r_{\text{m}}\leq 0.5$, we obtain
\begin{equation}
T\sim 0.5 m_{Pl}^{1/2}M_{\phi}^{1/2}.
\end{equation}
Hence, for the maximum temperature of the reheating period we find
\begin{equation}
T_{\text{reh}}\sim 3.29\times 10^{7}{\rm \ GeV}.
\end{equation}

\paragraph{Decay width  $\Gamma_{\phi}$ of the inflationary scalar field}

The relation between decay width of inflationary scalar field and $M_{\phi}$ is given by \cite{kolb2003reheating}
\begin{equation}\label{82}
\Gamma_{\phi}=\alpha _{\phi}M_{\phi}\sqrt{1-\left( \frac{T}{M_{\phi}}\right) ^2}\simeq \alpha _{\phi}M_{\phi}=\alpha _{\phi}\times 1{\rm \ keV},
\end{equation}
where $T$ is the temperature of the decay product, and $\alpha _{\phi}$ is a constant. As we have previously seen, the maximum temperature during reheating is of the order of $10^7$ GeV. The mass scale of the inflationary scalar field is of the order of the Planck Mass, which leads to $T\ll M_{\phi}$, so the decay width of inflationary scalar field is similarly proportional to  $M_{\phi}$. Hence for the decay width in the reheating model we adopt the value $\Gamma_{\phi}\approx \alpha _{\phi} \times 1{\rm \ keV}$.

\paragraph{Mass of the dark matter particle}\label{mdm}
The mass of dark matter particles is investigated model independently in \cite{de2010model}. Two kinds of dark matter particles, and their decoupling temperature from the dark matter galaxies, were obtained. One is possibly the sterile neutrino, gravitino, the light neutralino or majoron, with mass about $1\sim 2$ keV and decoupling temperature above $100$ GeV. The other one are the Weakly Interacting Massive particles (WIMP), with mass $\sim 100$ GeV and decoupling temperature $\sim 5$ GeV. From the analysis of reheating temperature above, we adopt the value of the mass of the dark matter particles as $m_{\text{DM}}\sim 1{\rm \ keV}$.

\paragraph{Other parameters}

For the reduce Planck mass we adopt the value $m_{Pl}=2.435\times 10^{18}$ GeV. If not otherwise specified, the numerical value of the spectral index is taken as $n_s=0.968$. In order to numerically solve the system of the reheating equations we use the initial conditions $a(0)=0.33$, $r_{\text{DM}}(0)=r_{\text{m}}(0)=n_{\text{m}}(0)=0$, $N_{\phi (0)}=N_{(\phi)0}$, $\phi (0)=\phi _0$, and $\left.\left(d\phi /dt\right)\right|_{\tau =0}=u_0$, respectively.

\section{Coherent scalar field model}\label{sect3}

As a first example of the analysis of the reheating process in the framework of the thermodynamics of open systems we assume that the inflationary scalar field can be regarded as a homogeneous oscillating coherent wave, with the energy density of the wave $\phi(t)$ given by \cite{B3, op8}
\begin{equation}\label{coherentphi}
\rho_{\phi}=m_{\phi}n_{\phi},
\end{equation}
where $m_{\phi}$ is the mass of the scalar field particle (inflaton). Then from relations (\ref{coherentphi}) and (\ref{relarhon}) we obtain
\begin{equation}
\rho_{\phi}=\dot{\phi}^2,
p_{\phi}=0,
\end{equation}
giving
\begin{equation}
V(\phi)=\frac{1}{2}\dot{\phi}^2.
\end{equation}
 In the following we assume that the energy density of the newly created matter and dark matter particles are small enough, and they do not affect the evolution of the energy density of the inflationary scalar field. That is, we assume the approximations
\begin{equation}
\rho_{\text{DM}},\rho_{\text{m}}\ll\rho_{\phi},
n_{\text{DM}},n_{\text{m}}\ll n_{\phi},
p_{\text{DM}},p_{\text{m}}\ll p_{\phi}.
\end{equation}

Then Eqs.~(\ref{geneqset})-(\ref{geneqsetf}) take the simplified form
\begin{equation}\label{simeqset}
2\dot{H}+3H^2=\Gamma_{\phi}m_{\phi}H,
\ee
\be\label{simeqsetf}
\dot{n}_{\text{DM}}+3Hn_{\text{DM}}=3\Gamma_{\text{DM}}H^2,\quad
\dot{n}_{\text{m}}+3Hn_{\text{m}}=3\Gamma_{\text{m}}H^2.
\end{equation}
Eqs.~ (\ref{simeqset})-(\ref{simeqsetf}) can be transformed into dimensionless form by introducing the dimensionless variables $\tau$, $h$, $N_{\text{DM}}$ and $N_{\text{m}}$, defined as
\bea
&&t=\frac{2}{\Gamma_{\phi}m_{\phi}}\tau, H=\frac{\Gamma_{\phi}m_{\phi}}{3}h,\nonumber\\
&&\hspace{-0.5cm}n_{\text{DM}}=\frac{2\Gamma_{\phi}m_{\phi}\Gamma_{\text{DM}}}{3}N_{\text{DM}}, n_{\text{m}}=\frac{2\Gamma_{\phi}m_{\phi}\Gamma_{\text{m}}}{3}N_{\text{m}}.
\eea
Hence we obtain the dimensionless equations
\be
\frac{{\rm d}h}{{\rm d}\tau}+h^2+h=0,
\ee
\be
\frac{{\rm d}N_{\text{DM}}}{{\rm d}\tau}+2hN_{\text{DM}}=h^2,\quad
\frac{{\rm d}N_{\text{m}}}{{\rm d}\tau}+2hN_{\text{m}}=h^2.
\end{equation}
The solution for $h$ is obtained as
\begin{equation}\label{simh}
h(\tau)=\frac{1}{e^{\tau}-1},
\end{equation}
while the time variations of the particle numbers are
\begin{equation}\label{simn}
N_{a}=\left( \frac{e^{\tau_0}-1}{e^{\tau}-1}\right)^2 N_{a0} e^{2(\tau-\tau_0)}+\frac{e^{2(\tau-\tau_0)}-1}{2(e^{\tau}-1)^2}, a={\rm DM}, {\rm m},
\end{equation}
where for dark matter $N_{a0}=N_{\text{DM0}}=N_{\text{DM}}(\tau_0)$, and $N_{a0}=N_{\text{m0}}=N_{\text{m}}(\tau_0)$ for the ordinary matter.

The scale factor in this model is given by,
\begin{equation}\label{sima}
a(\tau)=a_0 \left( \frac{e^{\tau}-1}{e^{\tau}}\right)^{\frac{2}{3}},
\end{equation}
where $a_0$ is an integration constant, while the deceleration parameter $q$ is found to be
\begin{equation}\label{simq}
q=\frac{3}{2}e^{\tau}-1.
\end{equation}
Since $q>0$ for all $\tau \geq 0$, it follows that the expansion rate of the universe continues to decrease. Inflation ends at the beginning of the reheating period.

Lets' consider now the beginning of the inflationary reheating, when $\tau$ is small. Then the solutions (\ref{simh})-(\ref{sima}) can be  approximated as
\begin{equation}
h\simeq \frac{1}{\tau},
a\simeq a_0\tau^{\frac{2}{3}},
\rho_{\phi}\simeq \frac{1}{\tau^2},
\ee
\be
N_{a}\simeq \frac{\tau-\tau_0+N_{a0}\tau_0^2}{\tau^2}, a={\rm DM}, {\rm m}.
\end{equation}
When $\tau$ reaches the value $\tau_{\text{max}}=2\tau_0-2N_0\tau_0^2$, the particle number reaches its maximum value,
\begin{equation}
n_{\text{(max)}a}=\frac{4\Gamma^2_a}{12\Gamma _at_0-9n_{a0}t_0^2},a={\rm DM}, {\rm m},
\end{equation}
where $n_{a0}=2\Gamma_{\phi}m_{\phi}\Gamma_a/3$. For dark matter, $\Gamma _a=\Gamma_{\text{DM}}$, and for matter $\Gamma _a=\Gamma_{\text{m}}$.

The entropy production of matter can also be calculated \cite{op8}, and for small times it shows a linear increase, so that
\begin{equation}
\frac{\mathcal{S}_{\text{m}}(t)}{S_{\text{m}}(t_0)}=\frac{\tau-\tau_0}{N_0\tau_0^2}+1.
\end{equation}

\section{Effect of the scalar field potential on the reheating dynamics}\label{sect4}

In the present Section we will investigate the reheating and generation of the matter content of the Universe, as described by the thermodynamics of open systems. In order to perform this study we will investigate the role played by the different scalar field potentials on the reheating dynamics.

\subsection{Large-field polynomial and linear potentials}

We will begin our analysis by assuming that the scalar field potential is a large-field polynomial potential, of the form
\begin{equation}
V(\phi)=M_{\phi}^2\left( \frac{\phi}{\mu}\right) ^p,
\end{equation}
where $p\geq 1$, $M_{\phi}$ is a parameter of the dimension of mass, while $\mu$ is a dimensionless constant. This kind of large-field polynomial potential is used in chaotic inflation models. A special case of the large-field polynomial potentials are the linear potentials, corresponding to $p=1$. These types of potentials are usually suggested by particle physics models, with dynamical symmetry breaking or nonperturbative
effects \cite{P1,P2}.  The inverse power law
potentials  appear in a globally supersymmetric $SU\left(N_c\right)$ gauge theory with $N_c$
colors and the condensation of $N_f$ flavors. Positive power-law potentials can drive an inflationary expansion, with the scalar field becoming progressively less
important as the cosmological evolution proceeds.  Hence the use of power law scalar field potentials can
convincingly justify the neglect of the scalar field terms during the post-inflationary and late times cosmological evolution in the Friedmann
equations.

The dimensionless energy density and  pressure of the inflationary scalar field take then the form
\begin{equation}
r_{\phi}=\frac{1}{2}\left(\frac{d\phi}{d\tau}\right)^2+\left( \frac{\phi}{\mu}\right) ^p,
P_{\phi}=\frac{1}{2}\left(\frac{d\phi}{d\tau}\right)^2-\left( \frac{\phi}{\mu}\right) ^p.
\end{equation}

The dimensionless equations Eqs.~(\ref{d1})-(\ref{d5}) describing the evolution of the Universe during the reheating phase can then be written as
\begin{equation}
\frac{d\phi }{d\tau }=u,  \label{l1}
\end{equation}
\bea
\frac{du}{d\tau }&=&-\frac{p}{\mu }\left( \frac{\phi }{\mu }\right) ^{p-1}-%
\sqrt{3}\sqrt{\frac{u^{2}}{2}+\left( \frac{\phi }{\mu }\right)
^{p}+r_{\text{DM}}+r_{\text{m}}}u-\nonumber\\
&&\frac{\alpha _{\phi }}{N_{\phi }}\left[ \frac{u^{2}}{2}%
+\left( \frac{\phi }{\mu }\right) ^{p}\right] u,  \label{l2}
\eea
\bea
\frac{dN_{\phi }}{d\tau }&=&-\sqrt{3}\sqrt{\frac{u^{2}}{2}+\left( \frac{\phi }{%
\mu }\right) ^{p}+r_{\text{DM}}+r_{\text{m}}}N_{\phi }-\nonumber\\
&&\alpha _{\phi }\left[ \frac{u^{2}}{2%
}+\left( \frac{\phi }{\mu }\right) ^{p}\right] ,  \label{l3}
\eea
\bea
\frac{dr_{DM}}{d\tau }&=&-\frac{4}{\sqrt{3}}\sqrt{\frac{u^{2}}{2}+\left( \frac{\phi }{%
\mu }\right) ^{p}+r_{\text{DM}}+r_{\text{m}}}r_{\text{DM}}+\nonumber\\
&&\gamma _{\text{DM}}\left[ \frac{u^{2}}{2}%
+\left( \frac{\phi }{\mu }\right) ^{p}\right] ,  \label{l4}
\eea
\bea
\frac{dr_{m}}{d\tau }&=&-\frac{4}{\sqrt{3}}\sqrt{\frac{u^{2}}{2}+\left( \frac{%
\phi }{\mu }\right) ^{p}+r_{\text{DM}}+r_{\text{m}}}r_{\text{m}}+\nonumber\\
&&\gamma _{\text{m}}\left[ \frac{u^{2}}{2}%
+\left( \frac{\phi }{\mu }\right) ^{p}\right] .  \label{l5}
\eea

The system of Eqs. (\ref{l1})-(\ref{l5}) must be integrated with the initial
conditions $\phi (0)=\phi _{0}$, $u(0)=u_{0}$, $N_{\phi }\left( 0\right)
=N_{\phi 0}$, $r_{DM}(0)=0$, and $r_{m}(0)=0$, respectively.

\begin{figure*}[htb]
\centering
\includegraphics[width=8.5cm]{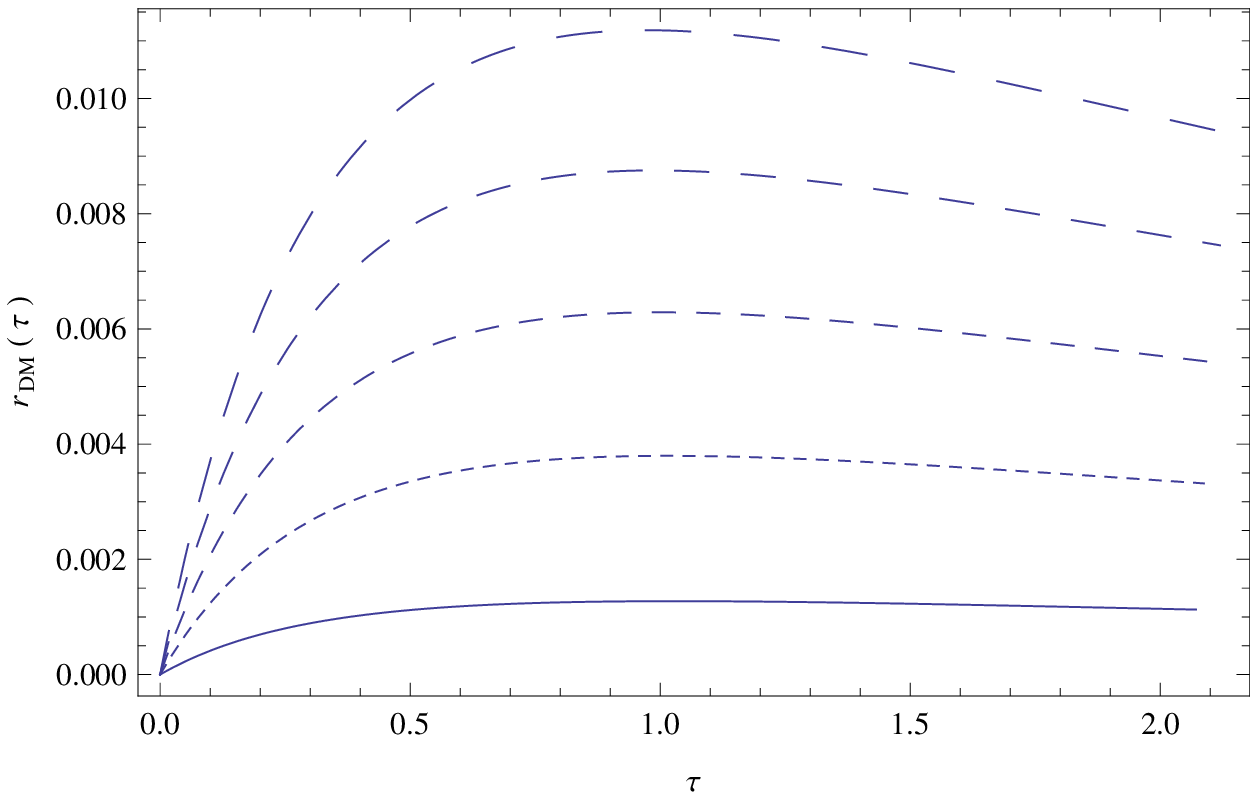}
\includegraphics[width=8.5cm]{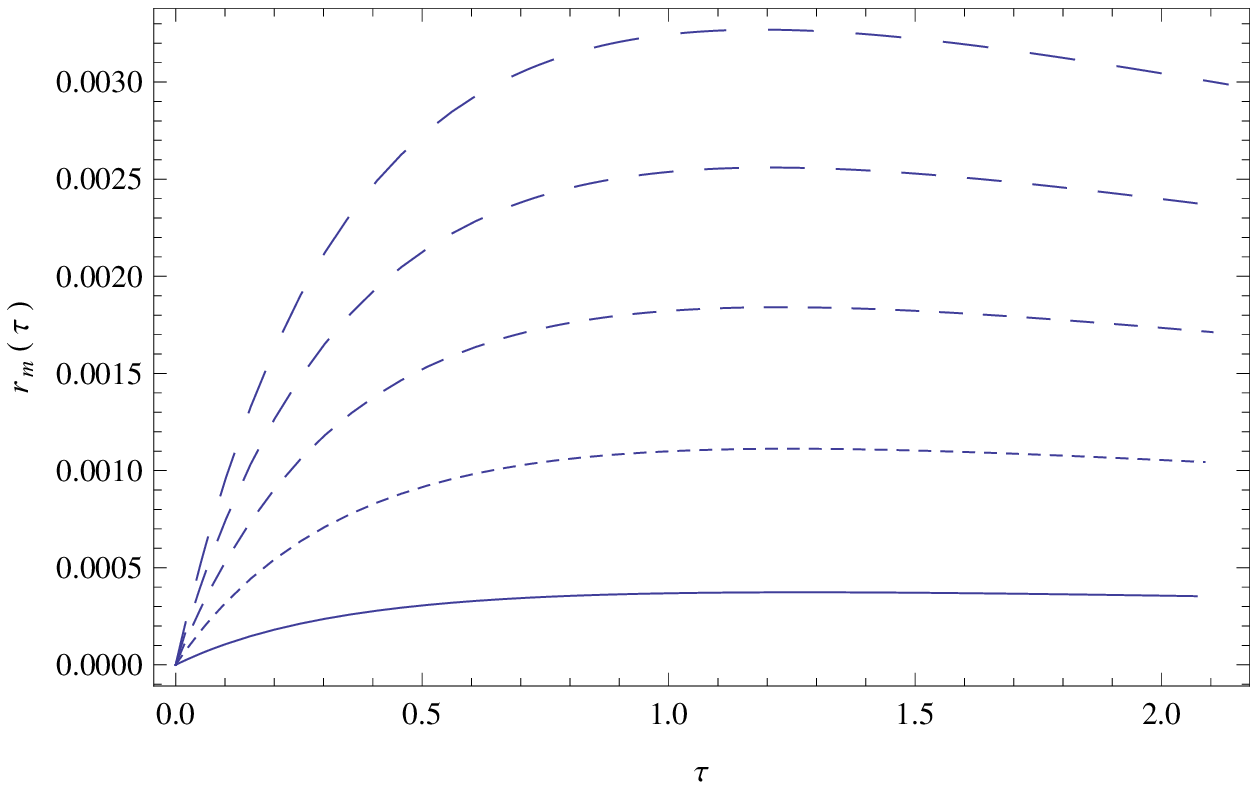}
\caption{Time variation of the dark matter energy density (left figure) and of the radiation (right figure) during the reheating period in the presence of a large field polynomial potential, for different values $\alpha _{\phi}$ of the decay width of the scalar field: $\alpha _{\phi}=0.01$ (solid curve), $\alpha _{\phi}=0.03$ (dotted curve), $\alpha _{\phi}=0.05$ (short dashed curve), $\alpha _{\phi}=0.07$ (dashed curve), and $\alpha _{\phi}=0.09$ (long dashed  curve), respectively. For $p$ we have adopted the value $p=1.8$, while $\mu =3.5$. The initial conditions used to numerically integrate Eqs. (\ref{l1})-(\ref{l5}) are $\phi (0)=1.1$, $u(0)=-1$, $N_{\phi }\left( 0\right)
=3$, $r_{DM}(0)=0$, and $r_{m}(0)=0$, respectively.}
\label{fig1}
\end{figure*}

\begin{figure*}[htb]
\centering
\includegraphics[width=8.5cm]{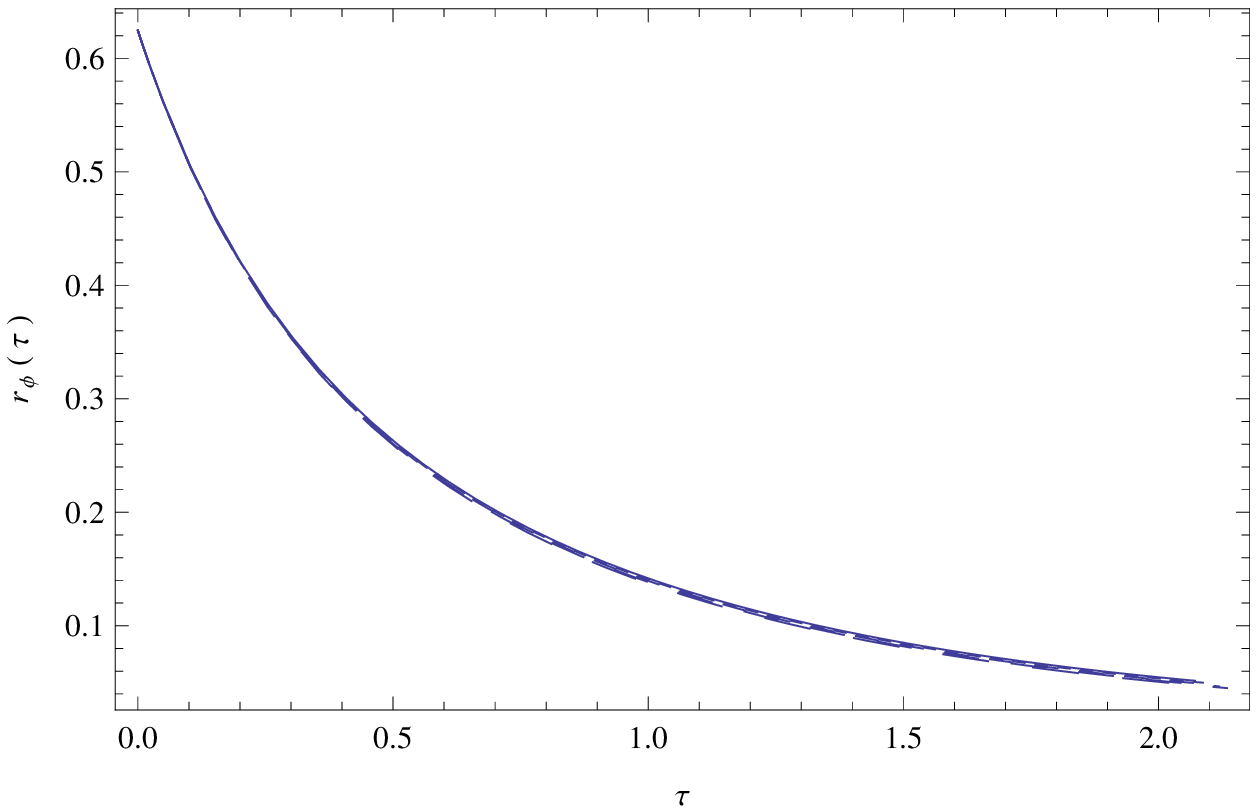}
\includegraphics[width=8.5cm]{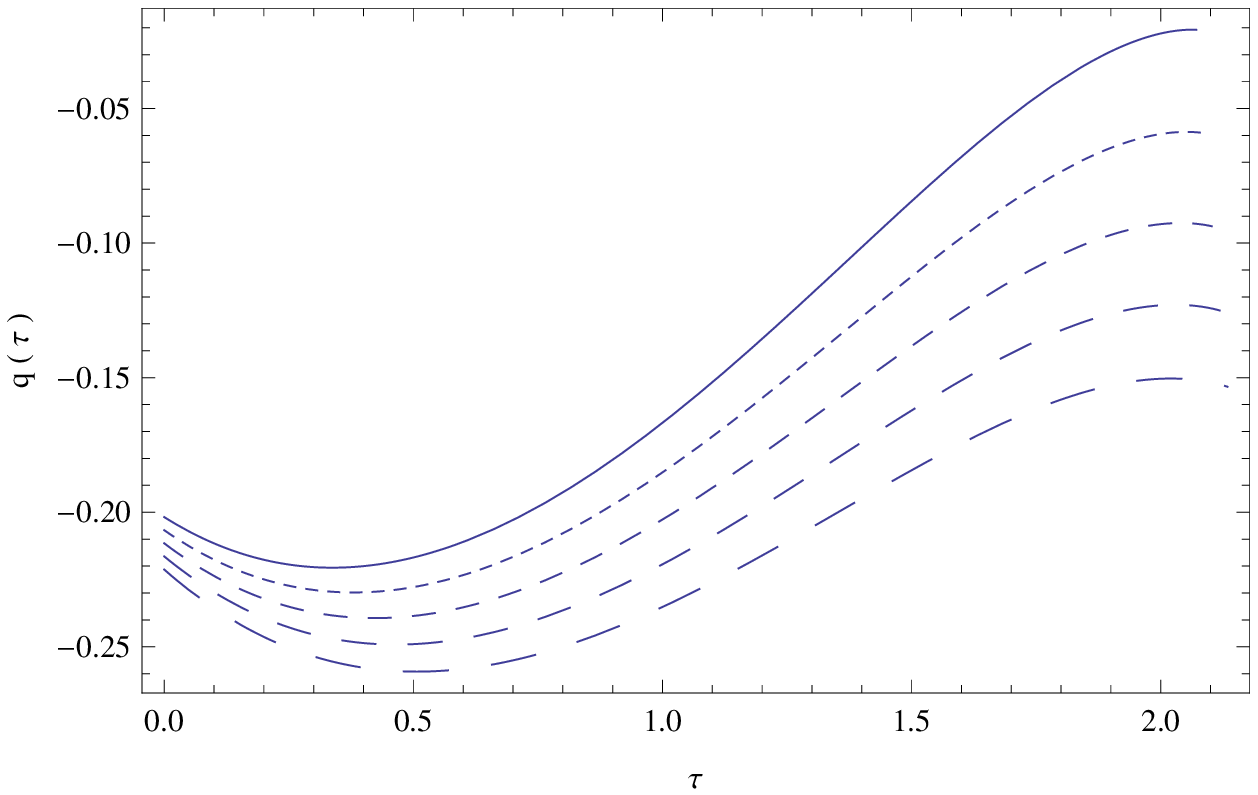}
\caption{Time variation of the energy density of the scalar field  (left figure) and of the deceleration parameter (right figure) during the reheating period in the presence of a large field polynomial potential, for different values $\alpha _{\phi}$ of the decay width of the scalar field: $\alpha _{\phi}=0.01$ (solid curve), $\alpha _{\phi}=0.03$ (dotted curve), $\alpha _{\phi}=0.05$ (short dashed curve), $\alpha _{\phi}=0.07$ (dashed curve), and $\alpha _{\phi}=0.09$ (long dashed  curve), respectively. For $p$ we have adopted the value $p=1.8$, while $\mu =3.5$. The initial conditions used to numerically integrate Eqs. (\ref{l1})-(\ref{l5}) are $\phi (0)=1.1$, $u(0)=-1$, $N_{\phi }\left( 0\right)
=3$, $r_{DM}(0)=0$, and $r_{m}(0)=0$, respectively. }
\label{fig2}
\end{figure*}

The variations with respect to the dimensionless time $\tau$ of the energy density of the dark matter, of the radiation, of the scalar field particle number, and of the Hubble function, are presented in Figs.~\ref{fig1} and \ref{fig2}, respectively. In order to numerically integrate Eqs.~(\ref{l1})-(\ref{l5}) we have fixed the parameters of the potential as $p=1.8$ and $\mu =3.5$, respectively. For the initial conditions of the scalar field and of its variation we have adopted the values $\phi (0)=1.1$, and $u(0)=-1.0$. The initial number of the scalar field particles was fixed to $N_{\phi }\left( 0\right)=3$, while for the coefficients $\gamma _{\text{DM}}$ and $\gamma _{\text{m}}$ we have used the values $\gamma _{\text{DM}}=(4/5)\alpha _{\phi}$ and $\gamma _{\text{m}}=(1/5)\alpha _{\phi}$.


As one can see from Figs.~\ref{fig1}, the evolution of the matter component (dark plus radiation) of the Universe can be divided into two distinct phases. In the first phase, the matter energy densities increase from zero to a maximum value $r_a^{(max)}$, $a={\rm DM}, {\rm m}$, which is reached at a time interval $\tau _{max}$, due to the energy transfer from the scalar field. For time intervals $\tau >\tau _{max}$, the expansion of the Universe becomes the dominant force describing the matter dynamics, leading to a decrease in the matter energy density. The energy density of the scalar field,  shown in the left panel of Fig.~\ref{fig2}, as well as the associated particle number decreases rapidly during the reheating process, and tends to zero in the large time limit. The energy density of the scalar field is not affected significantly by the variation of the numerical values of the parameter $\alpha _{\phi}$. The deceleration parameter of the Universe, depicted in the right panel of Fig.~\ref{fig2}, shows a complex evolution, depending on the nature of the particle creation processes. For the adopted initial values the particle creation phase in the inflationary history of the Universe begins with values of the deceleration parameter of the order of $q\approx -0.20$. During the particle creation processes the Universe continues to accelerate, a process specifically associate to the presence of the negative creation pressure, that induces a supplementary acceleration. This acceleration periods ends once the number of the newly created particles declines significantly, and at values of$q$ of the order of $q=-0.25$, the Universe starts decelerating, with the deceleration parameter tending to zero, corresponding to a marginally expanding Universe. When $\approx 0$, the energy density of the scalar field also becomes extremely small, and the Universe enters in the standard, matter dominated cosmological phase. with dark matter and radiation as the major components, plus a small remnant of the inflationary scalar field, which may play the role of the dark energy, and become again dominant in the latest stages of expansion.

We define the reheating temperature $T_{reh}$ as the temperature corresponding to the maximum value of the energy density of the relativistic matter $r_{{\rm m}}^{(max}$, whose energy density is defined as $\rho _{{\rm m}}=\left(\pi^2/30)\right)g_{reh}T^4$, where $g_{reh}\approx 100$ is the number of relativistic degrees of freedom. The maximum matter energy density $r_{{\rm m}}^{(max}$  is obtained from the condition $dr_{{\rm m}}/d\tau =0$,which gives the equation
\be\label{106}
4h\left(\tau_{max},T_{reh}\right)\frac{\pi ^2}{30}g_{reh}\frac{T_{reh}^4}{M_{\phi}^2m_{Pl}^2}=\frac{1}{5}\alpha _{\phi}r_{\phi}\left(\tau _{max}\right),
\ee
or, equivalently,
\be\label{107}
T_{reh}= 1.56\times 10^{9}\times \left[ \frac{24%
}{\pi ^{2}g_{reh}}\frac{\alpha _{\phi }r_{\phi }\left( \tau _{max
}\right) }{h\left( \tau _{max }\right) }\right] ^{1/4}\times M_{\phi
}^{1/2}\;{\rm GeV}.
\ee
Eqs.~(\ref{106}) and (\ref{107}) determine the reheating temperature as a function of $M_{\phi}$, the decay width of the scalar field, the value of the scalar field at maximum reheating, and its potential, respectively. If the reheating temperature is known from observations, from Eq.~(\ref{106}) one can constrain the parameters of the inflationary model, and of the scalar field potential. By adopting the bound $T_{reh}\leq 6\times 10^{10}$ \cite{Reh1}, we obtain the restriction on the parameter $M_{\phi}$ of the potential as
\be
M_{\phi}\leq 1.6\times 10^3 \times \left[ \frac{24%
}{\pi ^{2}g_{reh}}\frac{\alpha _{\phi }r_{\phi }\left( \tau _{max
}\right) }{h\left( \tau _{max }\right) }\right] ^{-1/4}\;{\rm GeV}.
\ee
The functions $r_{\phi}$ and $h$ are very little influenced by the modifications of the parameters of the potential or of $\alpha _{\phi}$ (but they depend essentially on the initial conditions). Hence for this class of potentials we approximate $r_{\phi }\left( \tau _{max
}\right)\approx 0.20$, and $h\left( \tau _{max}\right)\approx 0.70$. Hence we can approximate the coefficient in the scalar field parameter as
\be
\left[ \frac{24%
}{\pi ^{2}g_{reh}}\frac{\alpha _{\phi }r_{\phi }\left( \tau _{max
}\right) }{h\left( \tau _{max }\right) }\right] ^{-1/4}\approx 3.46\times \alpha _{\phi}^{-1/4}.
\ee
Therefore the second important parameter restricting the reheating temperature, or the potential parameters, is the decay width of the scalar field.  For $\alpha _{\phi}=10^{-3}$, $\alpha _{\phi}=10^{3/4}=5.62$, while for $\alpha _{\phi}=10^{3}$, $\alpha _{\phi}=10^{-3/4}=0.17$.
Hence the uncertainty in the knowledge of $\alpha _{\phi}$ can lead to variations over two orders of magnitude in the potential parameters.

\subsection{Small-field power law and small-field quadratic potentials }

The small-field polynomial potentials are of the form
\begin{equation}
V(\phi)=M_{\phi}^2\left[ 1-\left( \frac{\phi}{\mu}\right) ^p\right],
\end{equation}
where $\phi\ll\mu$ and $p\geq 2$. $M_{\phi}$ is a parameter of the dimension of mass, while $\mu$ is a dimensionless constant. These potentials appear as a result of a phase transition from spontaneous symmetry breaking. They are usually used in the "new inflation" models. A particular case of this class of potentials are small-field quadratic potentials, corresponding to $p=2$. Natural inflation models usually use a cosine potential, and the small-field quadratic potentials are its expansion near $\phi =0$ for $p=2$. For this class of potentials the dimensionless energy density and pressure of the scalar field takes the form
\bea
r_{\phi}&=&\frac{1}{2}\left(\frac{d\phi}{d\tau}\right)^2+\left[ 1-\left( \frac{\phi}{\mu}\right) ^p\right],\nonumber\\
P_{\phi}&=&\frac{1}{2}\left(\frac{d\phi}{d\tau}\right)^2-\left[ 1-\left( \frac{\phi}{\mu}\right) ^p\right].
\eea
The dimensionless equations describing particle production from inflationary scalar fields with small field polynomial potentials are
\begin{equation}\label{s1}
\frac{d\phi }{d\tau }=u,
\end{equation}
\bea\label{s2}
\frac{du}{d\tau }&=&\frac{p}{\mu }\left( \frac{\phi }{\mu }\right) ^{p-1}-\nonumber\\
&&\sqrt{3}\sqrt{\frac{u^{2}}{2}+\left[ 1-\left( \frac{\phi }{\mu }\right) ^{p}%
\right] +r_{DM}+r_{m}}u-\nonumber\\
&&\frac{\alpha _{\phi }}{N_{\phi }}\left[ \frac{u^{2}}{%
2}+1-\left( \frac{\phi }{\mu }\right) ^{p}\right] u,
\eea
\bea\label{s3}
\frac{dN_{\phi }}{d\tau }&=&-\sqrt{3}\sqrt{\frac{u^{2}}{2}+\left[
1-\left( \frac{\phi }{\mu }\right) ^{p}\right] +r_{DM}+r_{m}}N_{\phi
}-\nonumber\\
&&\alpha _{\phi }\left[ \frac{u^{2}}{2}+1-\left( \frac{\phi }{\mu }\right)
^{p}\right] ,
\eea
\bea\label{s4}
\frac{dr_{DM}}{d\tau }&=&-\sqrt{3}\sqrt{\frac{u^{2}}{2}+\left[ 1-\left( \frac{%
\phi }{\mu }\right) ^{p}\right] +r_{DM}+r_{m}}r_{M}+\nonumber\\
&&\alpha _{DM}\left[ \frac{%
u^{2}}{2}+1-\left( \frac{\phi }{\mu }\right) ^{p}\right] ,
\eea
\bea\label{s5}
\frac{dr_{m}}{d\tau }&=&-\frac{4}{\sqrt{3}}\sqrt{\frac{u^{2}}{2}+\left[
1-\left( \frac{\phi }{\mu }\right) ^{p}\right] +r_{DM}+r_{m}}r_{m}+\nonumber\\
&&\alpha
_{m}\left[ \frac{u^{2}}{2}+1-\left( \frac{\phi }{\mu }\right) ^{p}\right] .
\eea
The system of Eqs.~ (\ref{s1})-(\ref{s5}) must be integrated with the initial
conditions $\phi (0)=\phi _{0}$, $u(0)=u_{0}$, $N_{\phi }\left( 0\right)
=N_{\phi 0}$, $r_{DM}(0)=0$, and $r_{m}(0)=0$, respectively.

\begin{figure*}[htb]
\centering
\includegraphics[width=8.5cm]{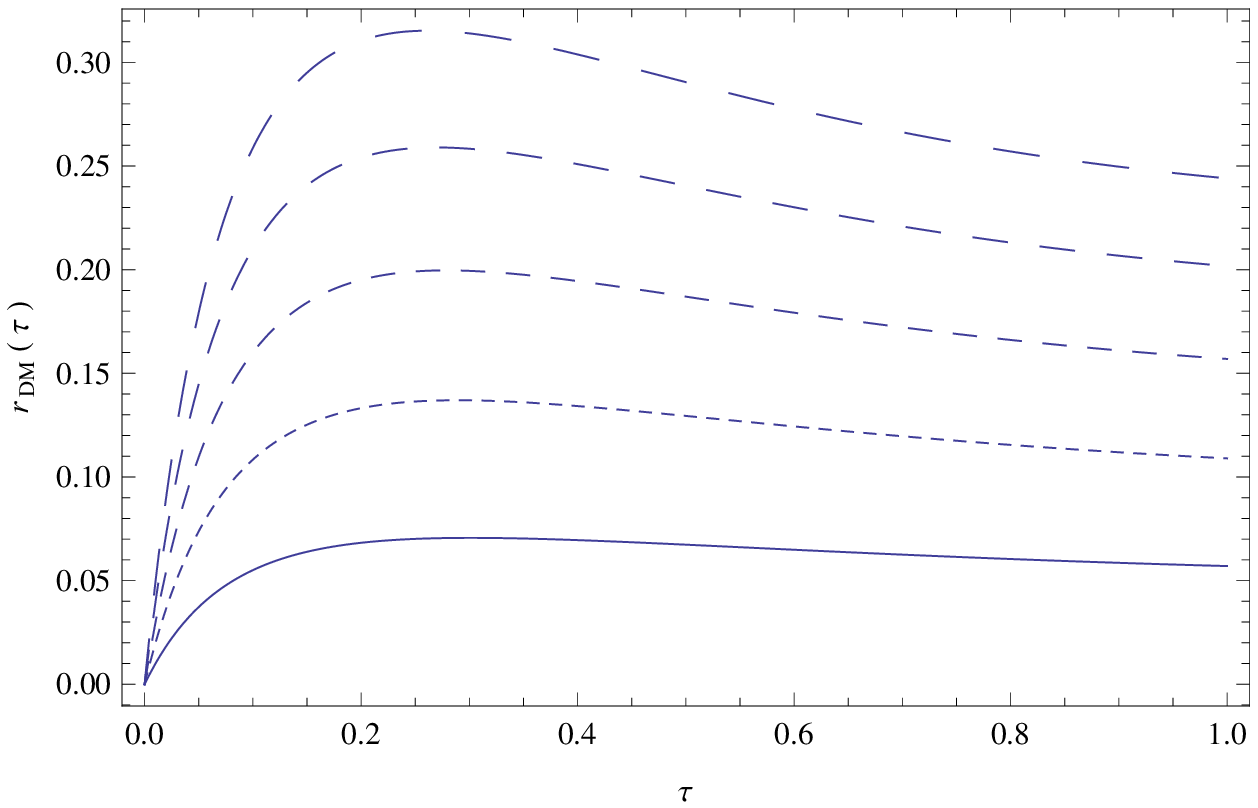}
\includegraphics[width=8.5cm]{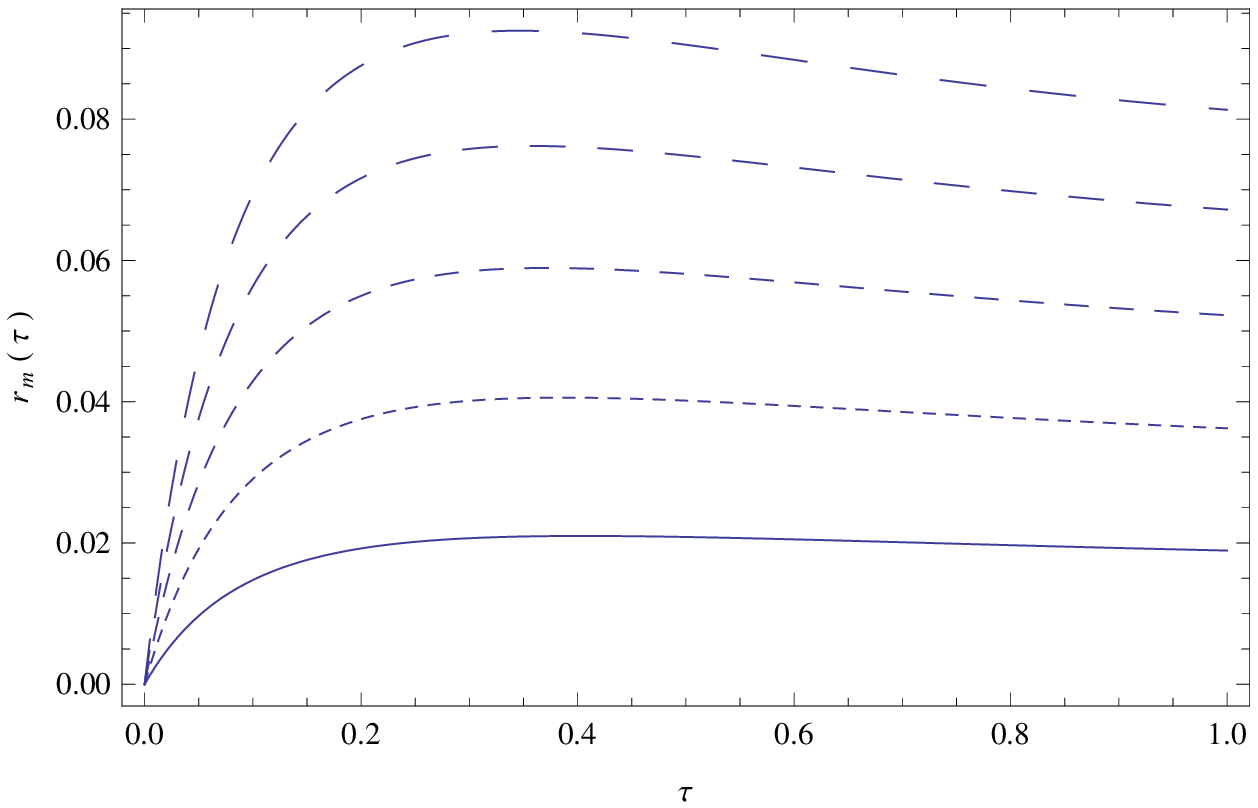}
\caption{Time variation of the dark matter energy density (left figure) and of the radiation (right figure) during the reheating period in the presence of a small field power law potential, for different values $\alpha _{\phi}$ of the decay width of the scalar field: $\alpha _{\phi}=0.15$ (solid curve), $\alpha _{\phi}=0.30$ (dotted curve), $\alpha _{\phi}=0.45$ (short dashed curve), $\alpha _{\phi}=0.60$ (dashed curve), and $\alpha _{\phi}=0.75$ (long dashed  curve), respectively. For $p$ we have adopted the value $p=3$, while $\mu =10^3$. The initial conditions used to numerically integrate Eqs. (\ref{s1})-(\ref{s5}) are $\phi (0)=1$, $u(0)=-4$, $N_{\phi }\left( 0\right)
=5$, $r_{DM}(0)=0$, and $r_{m}(0)=0$, respectively.}
\label{fig3}
\end{figure*}

\begin{figure*}[htb]
\centering
\includegraphics[width=8.5cm]{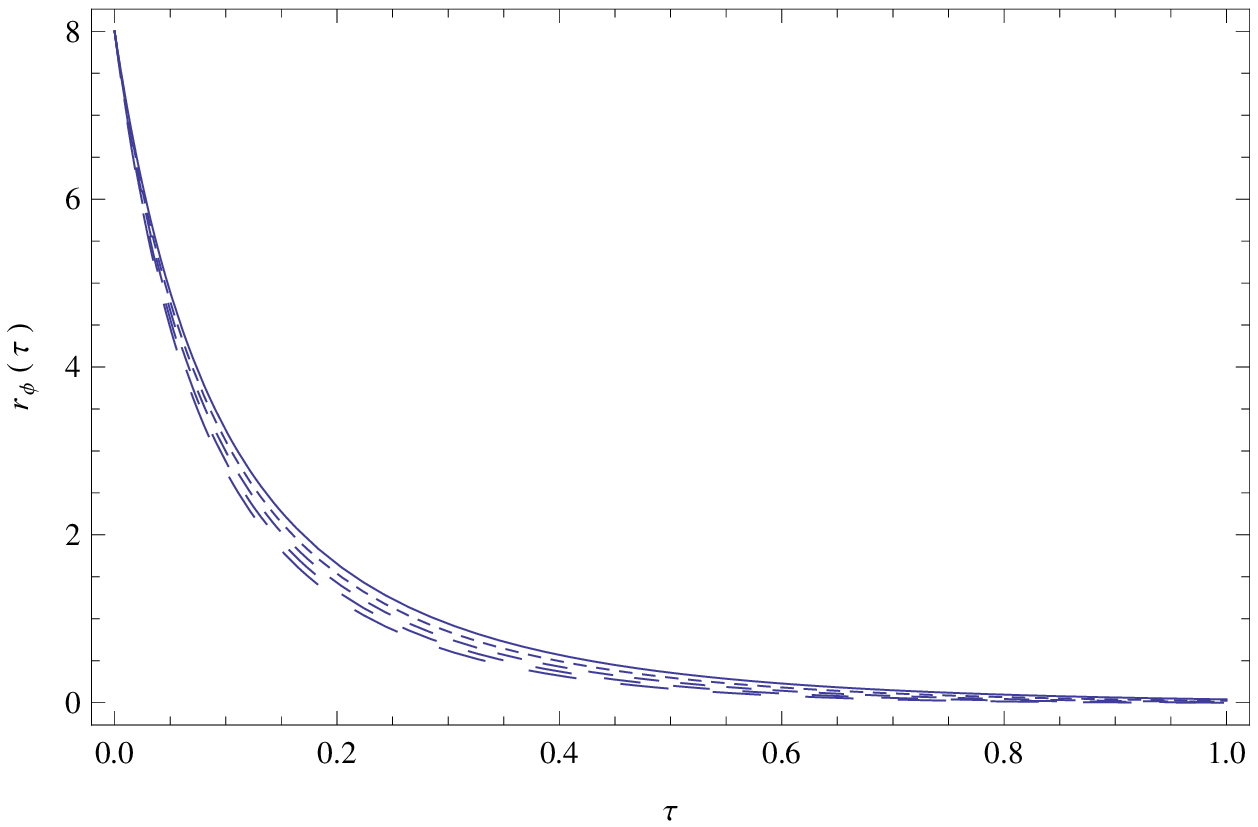}
\includegraphics[width=8.5cm]{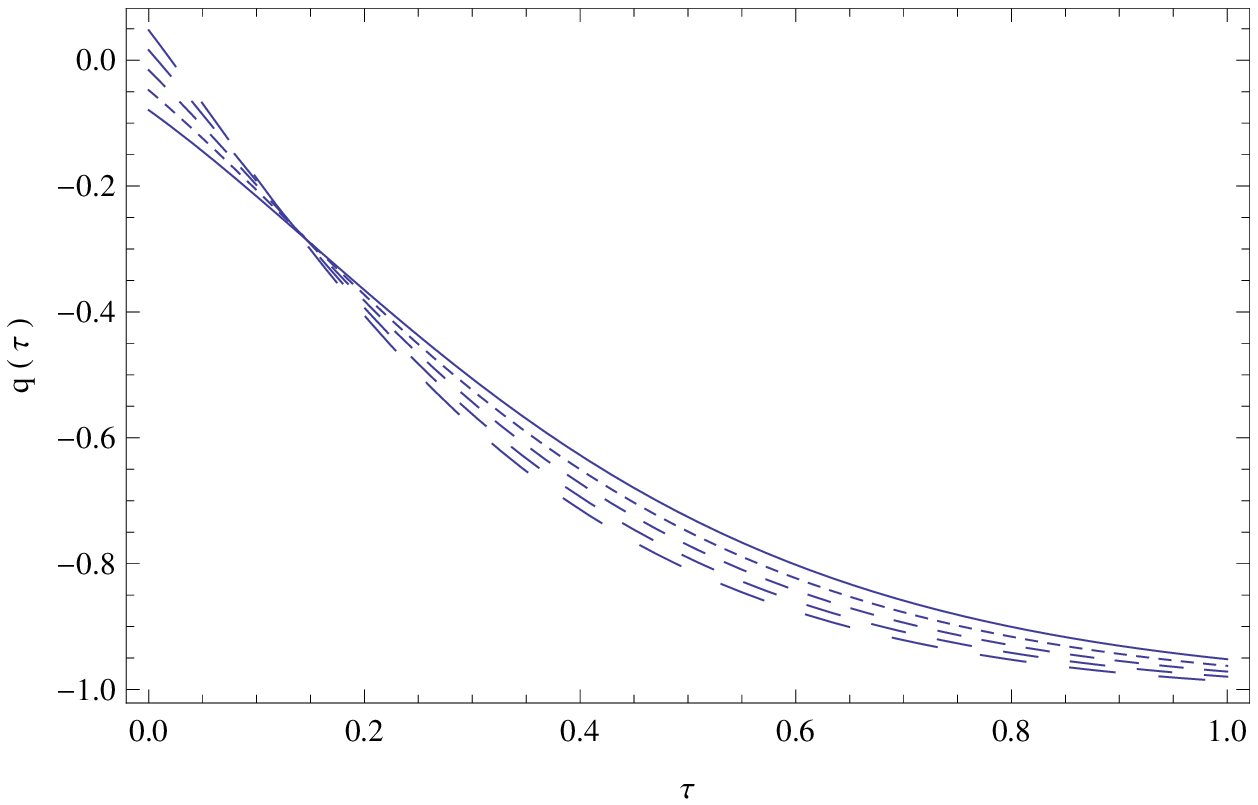}
\caption{Time variation of the energy density of the scalar field with  a small power law potential (left figure) and of the deceleration parameter (right figure) during the reheating period in the presence of irreversible matter creation , for different values $\alpha _{\phi}$ of the decay width of the scalar field: $\alpha _{\phi}=0.15$ (solid curve), $\alpha _{\phi}=0.30$ (dotted curve), $\alpha _{\phi}=0.45$ (short dashed curve), $\alpha _{\phi}=0.60$ (dashed curve), and $\alpha _{\phi}=0.75$ (long dashed  curve), respectively. For $p$ we have adopted the value $p=3$, while $\mu =10^3$. The initial conditions used to numerically integrate Eqs. (\ref{s1})-(\ref{s5}) are $\phi (0)=1$, $u(0)=-4$, $N_{\phi }\left( 0\right)
=5$, $r_{{\rm DM}}(0)=0$, and $r_{{\rm m}}(0)=0$, respectively.}
\label{fig4}
\end{figure*}

The variation of the dark matter energy density, of the radiation energy density, of the scalar field particle number, and of the Hubble function, respectively, are depicted in Figs.~\ref{fig3} and \ref{fig4}.


In order to  numerically integrate the system of Eqs.~(\ref{s1})-(\ref{s5}) we have fixed the value of $p$ at $p=3$, and the value of $\mu $ as $\mu =10^3$. We have slightly varied the numerical values of the parameter $\alpha _{\phi}$, which resulted in significant variations in the time evolution of the cosmological quantities. The initial conditions used for the numerical integration of the evolution equations are $u(0)=-4$, and $\phi (0)=1$, $N_{\phi}=5$, while the initial values of the energy density of the dark matter and of radiation were taken as zero. As one an see from the left panel of Fig.~\ref{fig3}, the energy density of the dark matter is initially a monotonically increasing function of time, reaching its maximum value $r_{{\rm DM}}$ at a time $\tau _{max}$. For time intervals $\tau >\tau _{max}$, the energy density of the dark matter decreases due to the expansion of the Universe. The time variation of the radiation, presented in the right panel of Fig.~\ref{fig3}, shows a similar behavior as the dark matter component. The energy density of the scalar field decreases, shown in the left panel of Fig.~\ref{fig4}, tends to zero in a finite time interval. The deceleration parameter of the Universe, presented in the right panel of Fig.~\ref{fig4}, shows an interesting behavior. For the chosen initial values the Universe is at $\tau =0$ in a marginally accelerating state, with $q\approx 0$. However, during the particle creation phase, the Universe starts to accelerate, and reaches a de Sitter phase with $q=-1$, at a time interval around $\tau =\tau _{fin}\approx 1$, when the energy density of the scalar field becomes negligibly small. Since $\rho _{\phi}\approx 0$ for $\tau >\tau _{fin}$, the particle creation processes basically stop, and the Universe enters in the matter dominated, decelerating phase. For the chosen range of initial conditions the numerical values of $\rho _{\phi}\left(\tau _{max}\right)$ and $h_{\tau_{\max}}$ are roughly the same as in the previous case, thus leading to the same restrictions on the potential parameter $M_{\phi}$.

\subsection{The exponential potential}

The exponential potentials are of the form
\begin{equation}
V(\phi)=M_{\phi}^2\exp{\left( \pm \sqrt{\frac{2}{p}}\phi\right) }
\end{equation}
where $M_{\phi}$ is a parameter of the dimension mass, and $p$ is a constant. Exponential potentials are sometimes used in extra dimensions models and superstring theories.  For example, an exponential potential is generated from compactification of the higher-dimensional supergravity or superstring theories in four-dimensional Kaluza-Klein type effective theories \cite{57}. The moduli fields associated with the geometry
of the extra-dimensions can have in string or Kaluza-Klein theories effective exponential type potentials, which are  due to the curvature of the internal spaces, or, alternatively, to the interaction on the internal spaces of the moduli fields with form fields. Non-perturbative effects such as gaugino condensation could also lead to the presence of exponential potentials \cite{58}.

The dimensionless energy density and pressure of the inflationary scalar field are
\be
r_{\phi}=\frac{1}{2}\left(\frac{d\phi}{d\tau}\right)^2+e^{ \pm\sqrt{\frac{2}{p}}\phi},
P_{\phi}=\frac{1}{2}\left(\frac{d\phi}{d\tau}\right)^2+e^{ \pm\sqrt{\frac{2}{p}}\phi}.
\ee

The dimensionless equations Eqs.~(\ref{d1})-(\ref{d5}) describing the evolution of the Universe in the presence of irreversible matter creation  can then be formulated as
\begin{equation}
\frac{d\phi }{d\tau }=u,  \label{e1}
\end{equation}
\bea\label{e2}
\frac{du}{d\tau }&=&\mp \sqrt{\frac{2}{p}}e^{\pm \sqrt{\frac{2}{p}}\phi }-%
\sqrt{3}\sqrt{\frac{u^{2}}{2}+e^{\pm \sqrt{\frac{2}{p}}\phi }+r_{DM}+r_{m}}u-\nonumber\\
&&\frac{\alpha _{\phi }}{N_{\phi }}\left[ \frac{u^{2}}{2}+e^{\pm \sqrt{\frac{2%
}{p}}\phi }\right] u,
\eea
\bea\label{e3}
\frac{dN_{\phi }}{d\tau }&=&-\sqrt{3}\sqrt{\frac{u^{2}}{2}+e^{\pm \sqrt{\frac{2%
}{p}}\phi }+r_{DM}+r_{m}}N_{\phi }-\nonumber\\
&&\alpha _{\phi }\left[ \frac{u^{2}}{2}%
+e^{\pm \sqrt{\frac{2}{p}}\phi }\right] ,
\eea
\bea\label{e4}
\frac{dr_{DM}}{d\tau }&=&-\sqrt{3}\sqrt{\frac{u^{2}}{2}+e^{\pm \sqrt{\frac{2}{p%
}}\phi }+r_{DM}+r_{m}}r_{M}+\nonumber\\
&&\alpha _{DM}\left[ \frac{u^{2}}{2}+e^{\pm \sqrt{%
\frac{2}{p}}\phi }\right] ,
\eea
\bea\label{e5}
\frac{dr_{m}}{d\tau }&=&-\frac{4}{\sqrt{3}}\sqrt{\frac{u^{2}}{2}+e^{\pm \sqrt{%
\frac{2}{p}}\phi }+r_{DM}+r_{m}}r_{m}+\nonumber\\
&&\alpha _{m}\left[ \frac{u^{2}}{2}+e^{\pm \sqrt{\frac{2}{p}}\phi }\right] .
\eea

The system of Eqs.~(\ref{e1})-(\ref{e5}) must be integrated with the initial
conditions $\phi (0)=\phi _{0}$, $u(0)=u_{0}$, $N_{\phi }\left( 0\right)
=N_{\phi 0}$, $r_{DM}(0)=0$, and $r_{m}(0)=0$, respectively.

\begin{figure*}[htb]
\centering
\includegraphics[width=8.5cm]{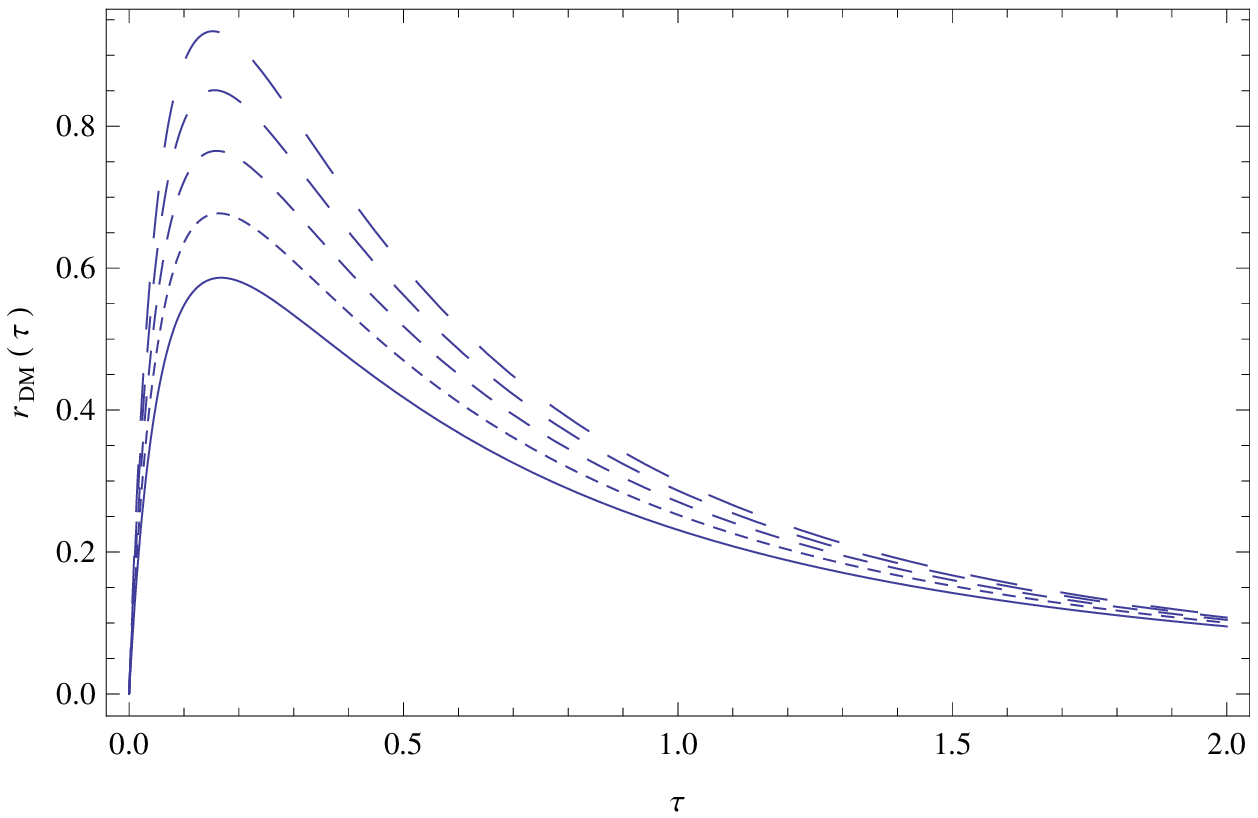}
\includegraphics[width=8.5cm]{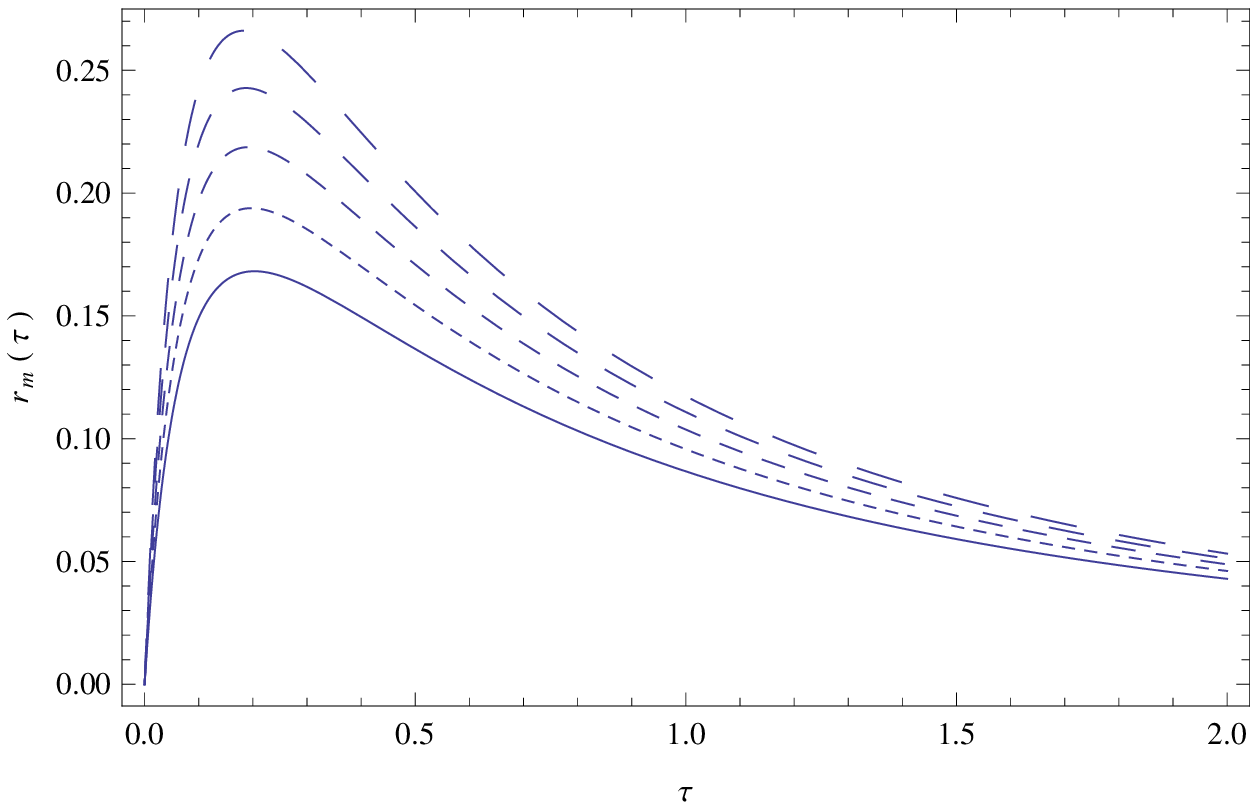}
\caption{Time variation of the dark matter energy density (left figure) and of the radiation (right figure) during the reheating period in the presence of an exponential scalar field potential, for different values $\alpha _{\phi}$ of the decay width of the scalar field: $\alpha _{\phi}=1.15$ (solid curve), $\alpha _{\phi}=1.35$ (dotted curve), $\alpha _{\phi}=1.55$ (short dashed curve), $\alpha _{\phi}=1.75$ (dashed curve), and $\alpha _{\phi}=1.95$ (long dashed  curve), respectively. For $p$ we have adopted the value $p=5$. The initial conditions used to numerically integrate Eqs. (\ref{e1})-(\ref{e5}) are $\phi (0)=15$, $u(0)=-5.5$, $N_{\phi }\left( 0\right)
=20$, $r_{{\rm DM}}(0)=0$, and $r_{{\rm m}}(0)=0$, respectively.}
\label{fig5}
\end{figure*}
\begin{figure*}[htb]
\centering
\includegraphics[width=8.5cm]{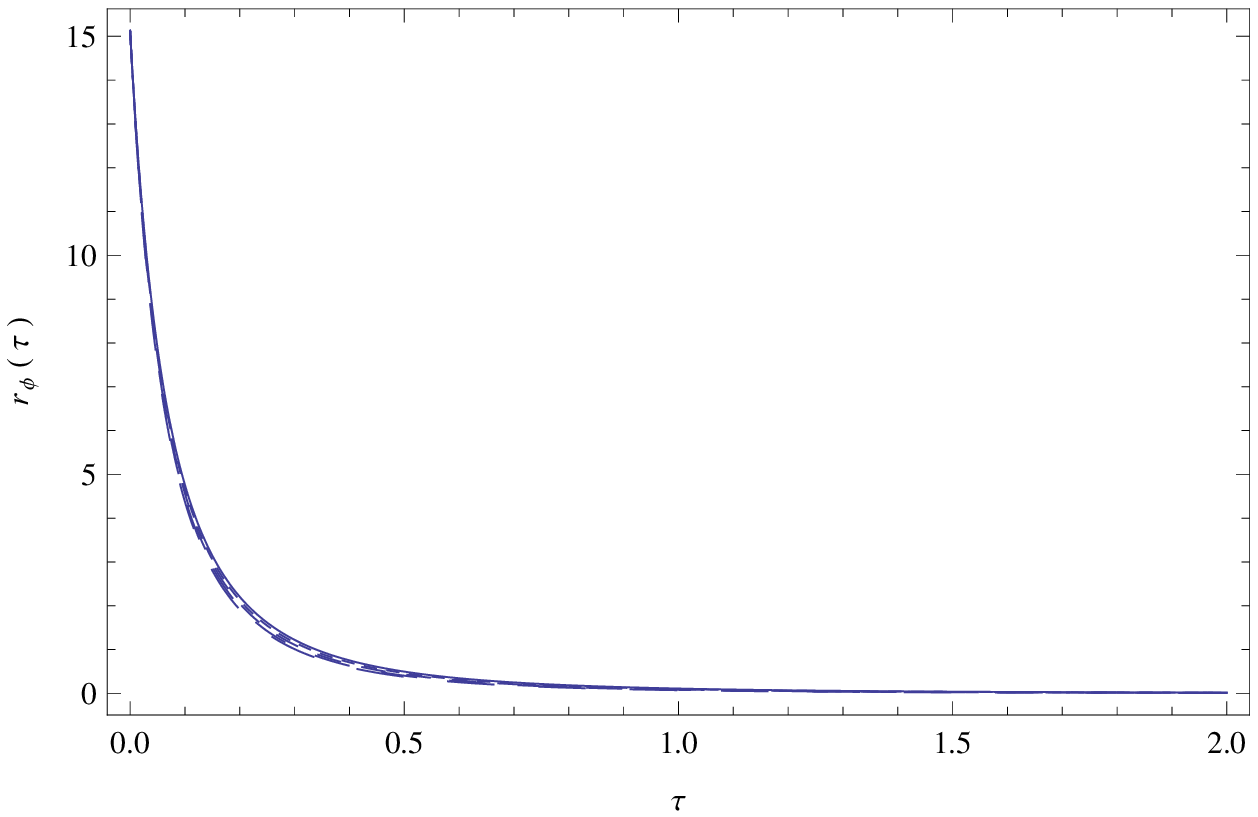}
\includegraphics[width=8.5cm]{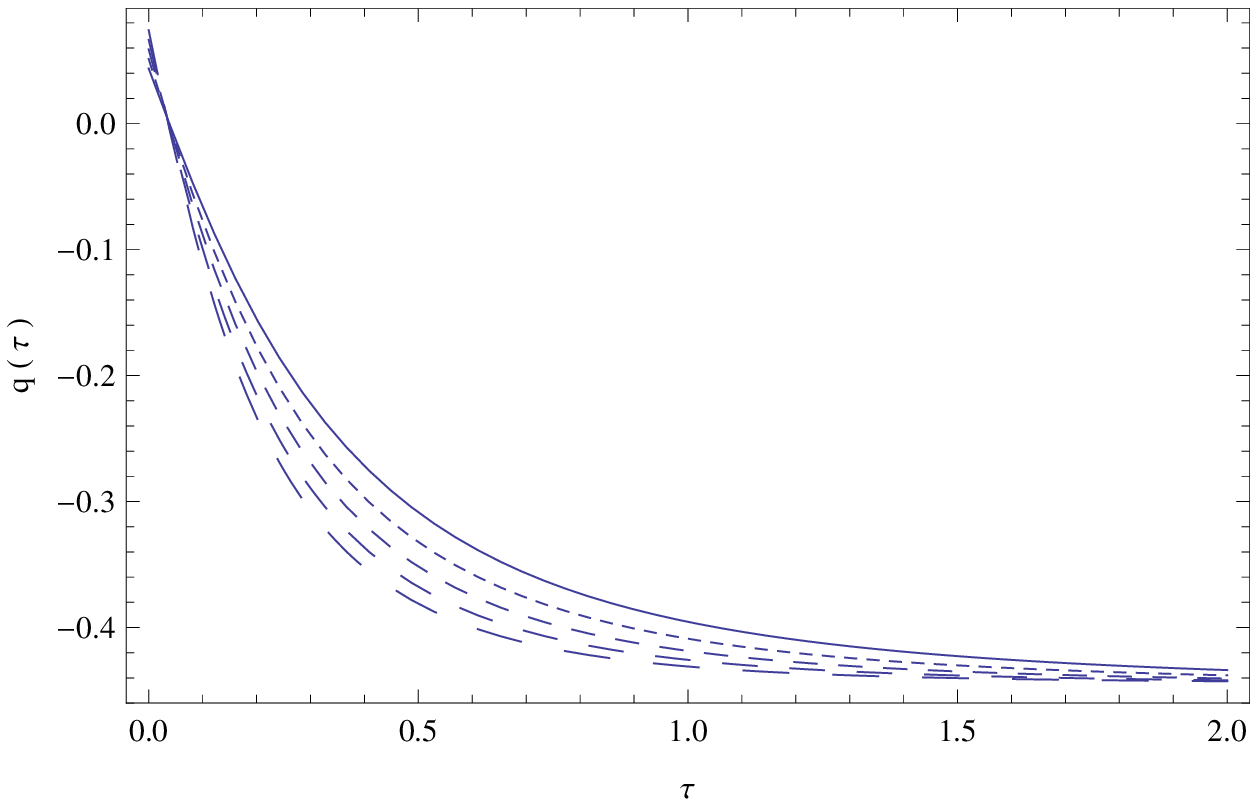}
\caption{Time variation of the energy density of the scalar field (left figure) and of the deceleration parameter (right figure) during the reheating period in the presence of an exponential scalar field potential, for different values $\alpha _{\phi}$ of the decay width of the scalar field: $\alpha _{\phi}=1.15$ (solid curve), $\alpha _{\phi}=1.35$ (dotted curve), $\alpha _{\phi}=1.55$ (short dashed curve), $\alpha _{\phi}=1.75$ (dashed curve), and $\alpha _{\phi}=1.95$ (long dashed  curve), respectively. For $p$ we have adopted the value $p=5$. The initial conditions used to numerically integrate Eqs. (\ref{e1})-(\ref{e5}) are $\phi (0)=15$, $u(0)=-5.5$, $N_{\phi }\left( 0\right)
=20$, $r_{{\rm DM}}(0)=0$, and $r_{{\rm m}}(0)=0$, respectively.}
\label{fig6}
\end{figure*}

In order to numerically integrate the cosmological evolution equations we have fixed the values of $p$ as $p=5$, and we have adopted the negative sign for the scalar field exponential potential. The initial conditions used for the numerical study are $\phi (0)=15$, $u(0)=-5.5$, $N_{\phi }\left( 0\right)
=20$, together with the null initial conditions for the matter energy densities. As one can see from the left panel of Fig.~\ref{fig5}, the amount of dark matter increases very rapidly during the first phase of the reheating. For the chosen values of the parameters and initial conditions, there is a sharp peak in the dark matter distribution, followed by a decreasing phase. A similar dynamic is observed for the time evolution of the radiation, shown in the right panel of Fig.~\ref{fig6}. The energy density of the scalar field, presented in the left panel of Fig.~\ref{fig6} decreases rapidly, and reaches the value $\rho _{\phi}\approx 0$ at $\tau \approx 1$. The evolution of the scalar field energy density is basically independent of the variation of the numerical values of the decay width of the scalar field $\alpha _{\phi}$. The numerical values of the deceleration parameter (right panel in Fig.~\ref{fig6}) do depend significantly on $\alpha _{\phi}$. At the beginning of the particle creation epoch the Universe is in a marginally inflating phase with $q\approx 0$. Particle creation induces a re-acceleration of the Universe, with the deceleration parameter reaching values of the order of $q\approx -0.15$ at the moment when the matter energy density reaches its peak value. However, the acceleration of the Universe continues, and at the moment the energy density of the scalar field becomes (approximately) zero, the deceleration parameter has reached values of the order of $q\approx -0.40$. Once the energy density of the scalar field is zero, the Universe starts its matter dominated expansion. For the chosen initial conditions $\rho _{\phi}\left(\tau _{max}\right)$ and $h_{\tau_{\max}}$ have significantly greater values than in the cases of the power law potentials. However, since the model parameters depend on their ratio, no significant variations in the reheating temperature or model parameters are expected.

\subsection{The Higgs Potential}

The Higgs potentials are of the form
\begin{equation}
V(\phi)=M_{\phi}^2\left[ 1-\left( \frac{\phi}{\mu}\right) ^2+\left( \frac{\phi}{\nu}\right) ^4\right]
\end{equation}
where $M_{\phi}$ is a parameter of the dimension mass,  and $\mu$ and $\nu$ are constants. The Higgs field is the only scalar field in the Standard Model of particle physics. The Higgs boson is a very special particle physics and in the Standard Model \cite{H1}. It provides a physical mechanism for the inclusion of the weakly interacting massive vector bosons in the Standard Model, and for giving masses to the fundamental particle of nature like quarks and leptons. The Higgs field may have also played an important role in cosmology.  It could have been the basic field making the Universe flat, homogeneous and isotropic, and it could have generated the small scale fluctuations that eventually led to cosmic structure formation. The Higgs field could have also allowed the occurrence of the radiation-dominated phase of the hot Big Bang \cite{H2}.

It is an intriguing question whether the initial scalar field driving the inflation is the same field as the Higgs field \cite{H2}. The dimensionless energy density and pressure of the inflationary Higgs scalar field are given by
\bea
\rho_{\phi}&=&\frac{1}{2}\left(\frac{d\phi}{d\tau}\right)^2+\left[ 1-\left( \frac{\phi}{\mu}\right) ^2+\left( \frac{\phi}{\nu}\right) ^4\right],  \nonumber\\
P_{\phi}&=&\frac{1}{2}\left(\frac{d\phi}{d\tau}\right)^2-\left[ 1-\left( \frac{\phi}{\mu}\right) ^2+\left( \frac{\phi}{\nu}\right) ^4\right].
\eea

The equations describing irreversible matter creation processes from a decaying scalar field in the presence of the Higgs type potential are
\begin{equation}
\frac{d\phi }{d\tau }=u,  \label{h1}
\end{equation}
\bea
\frac{du}{d\tau }&=&\phi \left( \frac{2}{\mu ^{2}}-\frac{4\phi ^{2}}{\nu ^{4}}%
\right) -\nonumber\\
&&\sqrt{3}\sqrt{\frac{u^{2}}{2}+1-\left( \frac{\phi }{\mu }\right)
^{2}+\left( \frac{\phi }{\nu }\right) ^{4}+r_{DM}+r_{m}}u-\nonumber\\
&&\frac{\alpha
_{\phi }}{N_{\phi }}\left[ \frac{u^{2}}{2}+1-\left( \frac{\phi }{\mu }%
\right) ^{2}+\left( \frac{\phi }{\nu }\right) ^{4}\right] u,  \label{h2}
\eea
\bea
\hspace{-0.5cm}&&\frac{dN_{\phi }}{d\tau }=-\sqrt{3}\times \nonumber\\
\hspace{-0.5cm}&&\sqrt{\frac{u^{2}}{2}+1-\left( \frac{\phi
}{\mu }\right) ^{2}+\left( \frac{\phi }{\nu }\right) ^{4}+r_{DM}+r_{m}}%
N_{\phi }-\nonumber\\
\hspace{-0.5cm}&&\alpha _{\phi }\left[ \frac{u^{2}}{2}+1-\left( \frac{\phi }{\mu }%
\right) ^{2}+\left( \frac{\phi }{\nu }\right) ^{4}\right] ,  \label{h3}
\eea
\bea
\hspace{-0.5cm}&&\frac{dr_{DM}}{d\tau }=-\sqrt{3}\times \nonumber\\
\hspace{-0.5cm}&&\sqrt{\frac{u^{2}}{2}+1-\left( \frac{\phi }{%
\mu }\right) ^{2}+\left( \frac{\phi }{\nu }\right) ^{4}+r_{DM}+r_{m}}%
r_{M}+\nonumber\\
\hspace{-0.5cm}&&\alpha _{DM}\left[ \frac{u^{2}}{2}+1-\left( \frac{\phi }{\mu }\right)
^{2}+\left( \frac{\phi }{\nu }\right) ^{4}\right] ,  \label{h4}
\eea
\bea
\hspace{-0.5cm}&&\frac{dr_{m}}{d\tau }=-\frac{4}{\sqrt{3}}
\times \nonumber\\
\hspace{-0.5cm}&&\sqrt{\frac{u^{2}}{2}+1-\left(
\frac{\phi }{\mu }\right) ^{2}+\left( \frac{\phi }{\nu }\right)
^{4}+r_{DM}+r_{m}}r_{m}+\nonumber\\
\hspace{-0.5cm}&&\alpha _{m}\left[ \frac{u^{2}}{2}+1-\left( \frac{%
\phi }{\mu }\right) ^{2}+\left( \frac{\phi }{\nu }\right) ^{4}\right] .
\label{h5}
\eea

\begin{figure*}[htb]
\centering
\includegraphics[width=8.5cm]{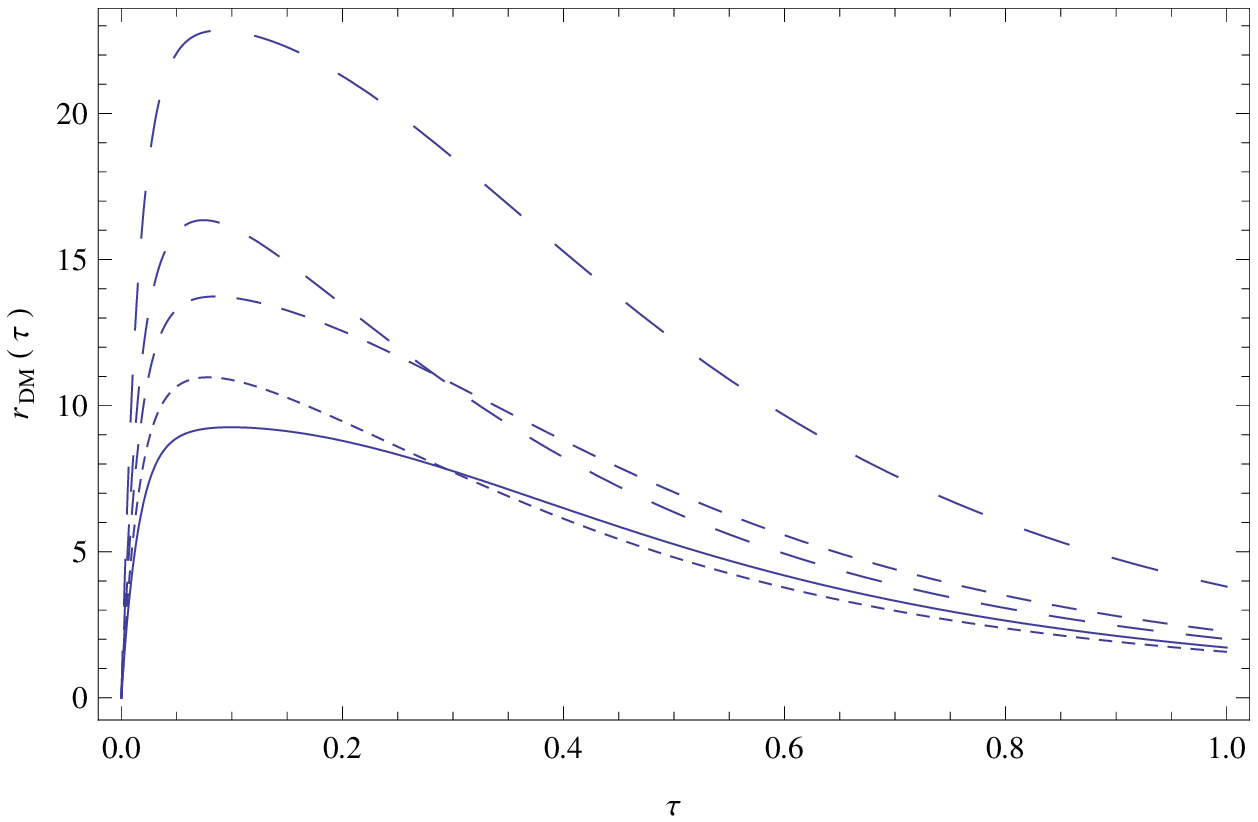}
\includegraphics[width=8.5cm]{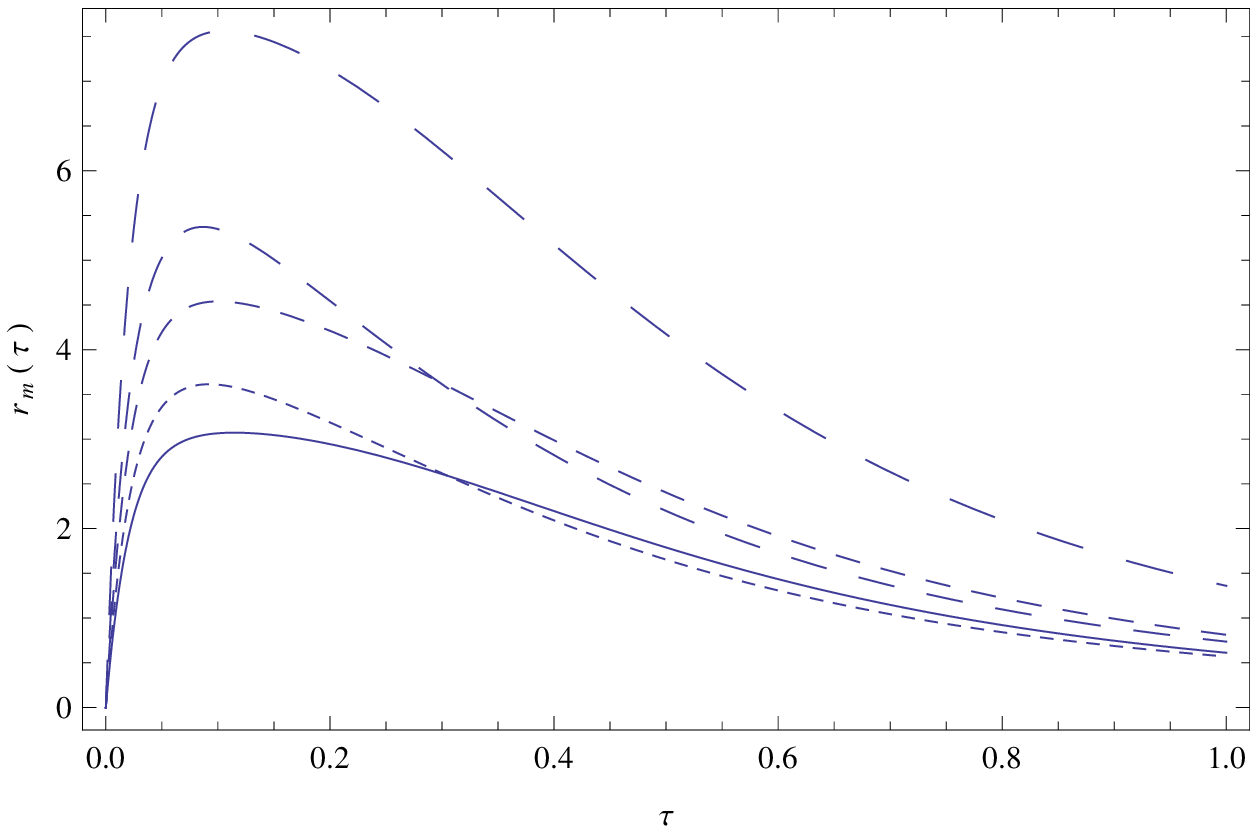}
\caption{Time variation of the dark matter energy density (left figure) and of the radiation (right figure) during the reheating period in the presence of a Higgs type scalar field potential, for different values $\alpha _{\phi}$ of the decay width of the scalar field: $\alpha _{\phi}=1.00$ (solid curve), $\alpha _{\phi}=1.20$ (dotted curve), $\alpha _{\phi}=1.40$ (short dashed curve), $\alpha _{\phi}=1.60$ (dashed curve), and $\alpha _{\phi}=2.50$ (long dashed  curve), respectively. For the potential parameters $\mu$  and $\nu $ we have adopted the value $\mu=2$ and $\nu =3$, respectively. The initial conditions used to numerically integrate Eqs.~(\ref{h1})-(\ref{h5}) are $\phi (0)=10.5$, $u(0)=-0.05$, $N_{\phi }\left( 0\right)
=100$, $r_{{\rm DM}}(0)=0$, and $r_{{\rm m}}(0)=0$, respectively.}
\label{fig7}
\end{figure*}

\begin{figure*}[htb]
\centering
\includegraphics[width=8.5cm]{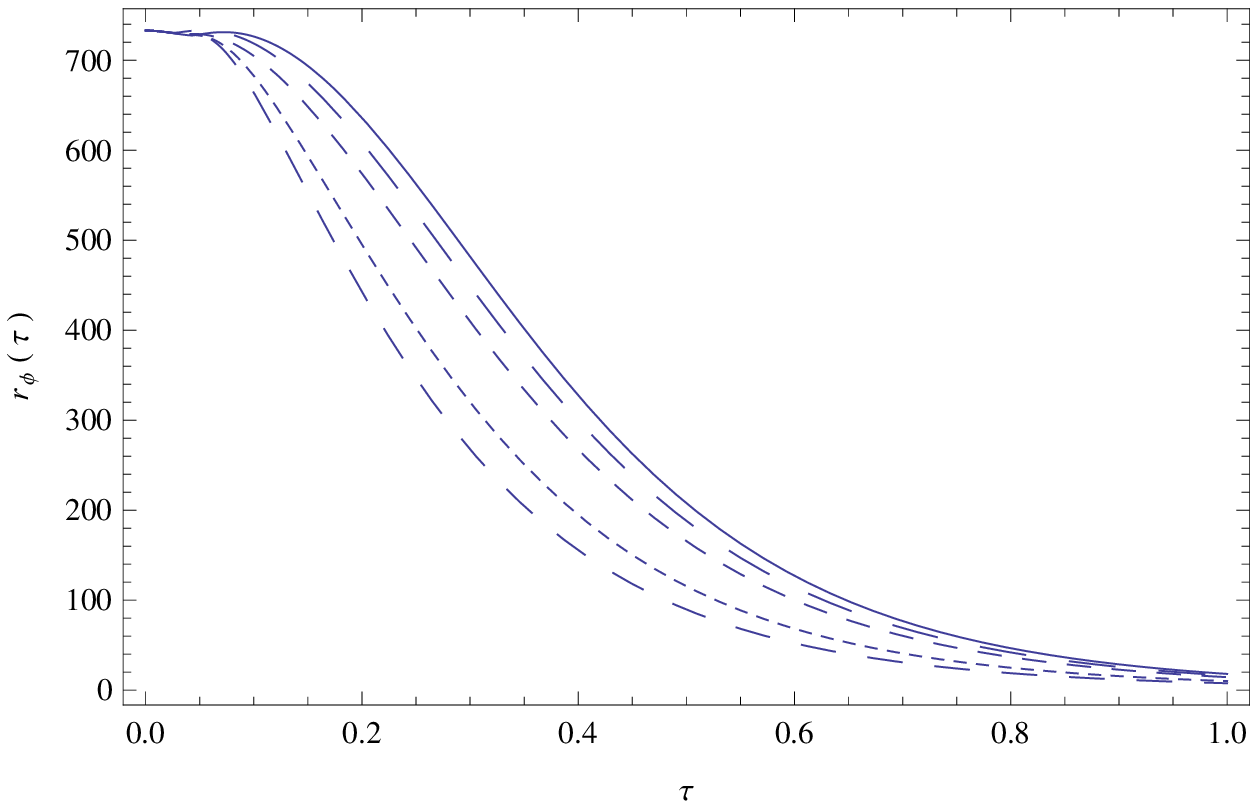}
\includegraphics[width=8.5cm]{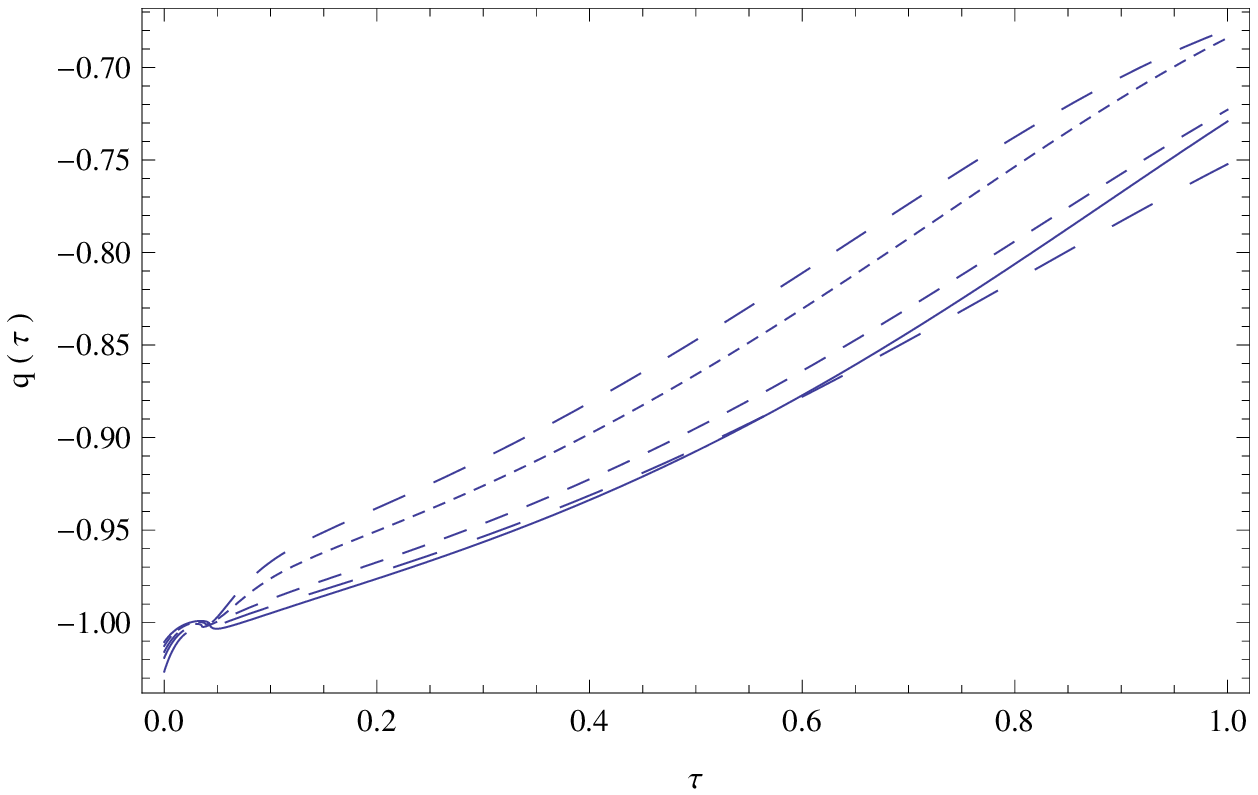}
\caption{Time variation of the energy density of the Higgs type scalar field  (left figure) and of the deceleration parameter (right figure) during the reheating period with irreversible particle creation for different values of $\alpha _{\phi}$: $\alpha _{\phi}=1.00$ (solid curve), $\alpha _{\phi}=1.20$ (dotted curve), $\alpha _{\phi}=1.40$ (short dashed curve), $\alpha _{\phi}=1.60$ (dashed curve), and $\alpha _{\phi}=1.80$ (long dashed  curve), respectively. For the potential parameters $\mu$  and $\nu $ we have adopted the value $\mu=2$ and $\nu =3$, respectively. The initial conditions used to numerically integrate Eqs.~(\ref{h1})-(\ref{h5}) are $\phi (0)=10.5$, $u(0)=-0.05$, $N_{\phi }\left( 0\right)
=100$, $r_{{\rm DM}}(0)=0$, and $r_{{\rm m}}(0)=0$, respectively.}
\label{fig8}
\end{figure*}

In order to obtain the numerical evolution of the Universe in the presence of irreversible matter creation triggered by the decay of a cosmological scalar field with Higgs type potential we fix the parameters of the potential as $\mu =2$ and $\nu =3$, respectively. For the initial conditions of the scalar field and of the scalar field particle number we have adopted the numerical values $phi (0)=10.5$ and $N_{\phi}(0)=100$, respectively, while for $u(0)$ we have taken the value $u(0)=-0.05$. The time variations of the dark matter and radiation energy densities, of the scalar field potential, and of the Hubble function are presented in Figs.~\ref{fig7} and \ref{fig8}, respectively.


For the adopted set of the parameters, the dark matter energy density, depicted in the right panel of Fig.~\ref{fig7}, increases from zero to a maximum value, and then begins to decrease, due to the combine effects of the expansion of the Universe, and of the decay of the scalar field. The radiation energy density, presented in the right panel of Fig.~\ref{fig7}, shows a similar behavior like dark matter. The evolution of both matter components is strongly dependent on the numerical values of the model parameters, as well as on the initial conditions. The energy density of the Higgs type scalar field, shown in the left panel of Fig.~\ref{fig8},  becomes negligibly small at $\tau \approx 1$. Its evolution also strongly depends on the model parameters. The deceleration parameter $q$, presented in the right panel of Fig.~\ref{fig8}, indicates that in the case of scalar field with Higgs potential the irreversible matter creation process starts when the Universe is in a de Sitter stage, with $q=-1$. During the first phase of the particle creation the Universe decelerates, but not significantly, and it continues its decelerating trend up to the moment when the energy density of the scalar field vanishes, at around $\tau \approx 1$, and standard matter dominated evolution takes over the expansionary dynamics of the Universe. However, at this moment in time the deceleration parameter still has a large negative value of the order of $q\approx -0.7$. For the chosen set of initial conditions the numerical values of $\rho _{\phi}\left(\tau _{max}\right)$ and $h_{\tau_{\max}}$ are similar to the case of the exponential potential, and hence no significant differences in the predicted reheating temperature or in model parameter constraints can take place.

\section{Discussions and final remarks}\label{sect5}

In the present paper we have investigated the implications of the irreversible thermodynamic description of particle creation during the reheating period of the early Universe. By interpreting the post-inflationary Universe as an open system with energy-matter transfer between the fluid components, a complete thermodynamic description of the basic physical processes taking place at the end of inflation can be obtained.  The basic feature of our model is the decay of the inflationary scalar field and the transfer of its energy to the newly created particles. As for the matter content of the post-inflationary Universe we have assumed that it consists of pressureless dark matter, and a relativistic, radiation type component. By using the fundamental principles of the thermodynamics of open systems and the Einstein field equations, a systematic and consistent approach describing the time evolution of the matter components after inflation can be obtained.   We have investigated the effects on particle creation of various forms of the scalar field potential, which generally  lead to similar qualitative results for the reheating scenario.  What we have found from our numerical analysis is that the choice of the functional form of the scalar field potential has an essential influence on the cosmological dynamics. This influence can be seen in the time interval for which the maximum energy densities of the dark matter and radiation are obtained, in the decay rate of the energy density of the scalar field, and its conversion to matter, as well as in the evolution of the deceleration parameter. Moreover, the cosmological dynamics also depends on the numerical values of the potential parameters.

However, the scalar field potentials have a unique quantitative imprint on the matter creation during reheating. By adopting a set of initial conditions for the physical variables,  the general equations describing reheating can be solved numerically.  During reheating, the particle number density of the scalar field drops to near zero, while the number of the dark matter and matter particles increases from zero to a maximum value, thus heating up the Universe to a maximum reheating temperature $T_{\text{reh}}$. The reheating temperature depends on the physical parameters of the inflation (scalar field energy density, field decay width, potential parameters), and on the cosmological quantities, like the value of the Hubble function at the moment when the matter density reached its maximum value. By using the general constraint on the reheating temperature as obtained in \cite{Reh1}, we have obtained some general constraints on the mass parameter in the scalar field potential. However, in our investigation these results are only of qualitative nature, since a significant increase in the accuracy of the observational data is still needed to obtain an accurate reconstruction of the physical conditions in the early Universe.  Large-field potential models lead to similar results as the kinetic dominated ones, since the observation of the scalar spectral index $n_s$ constrains the value of $p$ to be relatively small, with  observations preferring the potentials with $p=2$ \cite{Re5}.  The exponential potential models speed up the matter creation process, specifically the preheating process to $\tau<0.2$, and keeps creating new particles even after the matter energy density has reached its maximum value. The process is similar to the one induced by the Higgs potentials model. In our present approach we have neglected the backreaction of the newly created particles on the dynamics of the Universe.

We have also carefully investigated the behavior of the deceleration parameter during the reheating period. A general result is that for all considered potentials the Universe still accelerates during the reheating phase. This is not an unexpected result from a physical point of view, since in the formalism of the thermodynamics of open systems particle creation is associated with a negative creation pressure, that accelerates the system. Irreversible particle creation has been proposed as a possible explanation of the late time acceleration of the Universe \cite{a1,a2,a3,a4,a5,a6}. A similar process takes place in the irreversible thermodynamic description of the reheating, with the Universe acquiring a supplementary acceleration. Moreover, this acceleration extends beyond the moment of the reheating (when the maximum temperature is reached), and indicates that the Universe does not enter the standard, matter and radiation dominated phase, in a decelerating state. Hence the use of the thermodynamic of open systems with irreversible matter creation leads to an accelerating expansion of the Universe during the particle production stage, corresponding to the post-inflationary reheating. This effect is strongly scalar field potential dependent, and while for some models the Universe acceleration is marginal, with $q\approx 0$ (coherent scalar field and large field polynomial potential with $V(\phi)\propto \phi ^p$), for other types of potentials (small field power law, exponential, Higgs), the deceleration parameter at the end of the reheating period could have negative values in the range $q\in (-1,-0.4)$. These extremely  accelerating models may be unrealistic, since they would indeed pose some serious problems to the nucleosynthesis and structure formation, and contradict some well-established cosmological bounds on the chemical composition of the Universe, and the time frame the first chemical elements were produced.

On the other hand, in the present model the cosmological behavior and particle creation rates also strongly depend on the scalar field potential parameters, whose full range has not been systematically and completely investigated in the present study. Therefore the use of the thermodynamic of open processes for the description of the post-inflationary reheating may also provide some strong constraints on the admissible types of scalar field potentials. In this context we would like to note that model independent constraints on the cosmological parameters can be obtained within the framework of the  so-called cosmography of the Universe. Cosmography is essentially a Taylor expansion as applied to cosmology. Model independent constraints on the evolution of the deceleration parameter have been found by using the cosmographic approach in \cite{q1,q2,q3,q4,q5}.

When the temperature of the decay products $T$ approaches the mass scale $M_{\phi}$ in the scalar field potentials, $T\rightarrow M_{\phi}$, then, as one can see from Eq.~(\ref{82}), the decay width $\Gamma _{\phi}$ of the scalar field tends to zero, $\Gamma _{\phi}\rightarrow 0$. Moreover, the energy density of the scalar field becomes negligibly small, meaning that after the end of the reheating phase, corresponding to $\Gamma _{\phi}\approx 0$ and $\rho _{\phi}\approx 0$, the number of the dark matter particles, and of the photons, is conserved. Hence in the post reheating phase the cosmological evolution of the Universe is governed by the equations
\be\label{fin1}
3H^2=\frac{1}{m_{Pl}^2}\left(\sum _i{\rho _i}\frac{1}{2}\dot{\phi}^2+V(\phi)\right),
\ee
\be\label{fin2}
\dot{\rho}_i+3H\left(\rho _i+p_i\right)=0, i=1,2,..,
\ee
\be\label{fin3}
\ddot{\phi}+3H\dot{\phi}+V'(\phi)=0,
\ee
where $\rho _i$ and $p_i,i=1,2,..$ denotes the thermodynamic densities and pressures of the matter components (radiation, dark matter, baryonic matter etc.) of the Universe.  Eqs.~(\ref{fin1})-(\ref{fin3}) represents a cosmological model in which the Universe is composed from matter and dark energy, described by a scalar field with a self-interaction potential $V(\phi)$. In this way we have recovered the quintessence dark energy models,  in which dark energy is described by a canonical scalar field $\phi $ minimally coupled to gravity \cite{Tsu}. As compared to other scalar-field models, like phantom and k-essence dark energy scenarios, the quintessence model
is the simplest scalar-field approach not having major theoretical shortcomings such as the
presence of ghosts and of the Laplacian instabilities. Moreover, a slowly varying scalar field together with a potential
$V(\phi)$ can lead to the late-time acceleration of the Universe. The quintessence scalar field has
the important property of being very weakly coupled to matter (baryonic and dark),  but it contributes a negative pressure to the global equation of state.  The magnitude of the energy scale of the quintessence potential must be of the
order of $\rho _{DE}\approx  10^{-47}$  GeV$^4$ at the present time,  a value that is much smaller than that of the inflaton potential triggering the inflationary expansion. In order to achieve a late time cosmic acceleration, the mass $m_{\phi}$  of the quintessence field, given by
$m_{\phi}^2=d^2V (\phi)/d\phi ^2$ must be extremely small, of the order of  $\left|m_{\phi}\right| \sim  H_0 \approx  10^{−33}$ eV \cite{Tsu},
where $H_0$ is the present day value of the Hubble function. The evolution of the Universe as described by the Eqs.~(\ref{fin1})-(\ref{fin3}) starts from the initial conditions fixed at the end of reheating, including the initial matter densities, scalar field and Hubble function values.   In the early stages of the evolution of the Universe (post-reheating phase) the quintessence scalar field had a very small contribution to the energy balance of the Universe, but during the later cosmological evolution it decreases slower with the scale factor as compared to the matter and radiation densities,  and consequently it is dominant at the present time, triggering an exponential increase of the scale factor.  Therefore even the Universe enters in the conservative phase in an accelerating state, in the first stages of the conservative evolution, in which the effect of the scalar field can be neglected, the cosmological evolution will strongly decelerate. By assuming that the energy density of the scalar field is negligibly small, and the Universe is matter dominated, then the scale factor is given by $a(t)=t^{2/3}$, with the corresponding deceleration parameter having the value $q=1/2$.

An interesting and important question is the physical nature of the dark matter particles that could have been  created during reheating. Presently, the true physical nature of the dark matter particle is not known. One possibility is that dark matter may consist of ultra-light particles, with masses of the order of $m\approx 10^{-24}$ eV (see \cite{Lee} and references therein). Such an ultra-light dark matter  particle may represent, from a physical point of view, a  pseudo Nambu-Goldstone boson. Other important ultra-light dark matter particle candidates are the axions, with masses of the order of  $m\leq 10^{-22}$ eV \cite{Park}. Very low mass particles can be easily created even in very weak gravitational fields, and hence their production in large numbers during reheating could have taken place. Another hypothesis about the nature of dark matter is represented  by the so-called Scalar Field Dark Matter (SFDM) models \cite{Mat}. In these models dark matter is described as a real scalar field, minimally coupled to gravity, with the mass of the scalar field particle having a very small value, of the order of $m <10^{-21}$ eV. An important property of scalar field dark matter models is that at zero temperature all particles condense to the same quantum ground state, thus forming a Bose-Einstein Condensate (BEC). Hence, from a physical point of view it turns out that SFDM models are equivalent to the BEC dark matter models \cite{Bose1, Bose2, Bose3,Bose4, Bose5, Bose6}. On the other hand, in the open irreversible thermodynamic  model of reheating introduced in the present paper, dark matter particle creation can take place also in the form of a scalar field, one we assume that the scalar field does not fully decays into some ordinary matter.  Hence, scalar field dark matter could be a particular result of the inflationary  scalar field dynamics during the reheating phase in the evolution of the Universe.

Particle creation can also be related to what is called the arrow of time: a physical process that provides a direction to time, and distinguishes the future from the past. There are two different arrows of time: the thermodynamical arrow of time, the direction, in which entropy increases,  and the cosmological arrow of time, the direction in which the Universe is expanding. Particle creation introduces asymmetry in the
evolution of the Universe, and enables us to assign a thermodynamical arrow of time, which agrees, in the present  model, with the cosmological one. This coincidence is determined, in a natural way, by the physical nature of inflation.

The present day observations cannot accurately constrain the main predictions of the present model. However, even without a full knowledge of the true cosmological scenario one can make some predictions about reheating. In the models we are considering the energy density of the inflationary scalar field does not drop completely to zero during reheating. This may suggest that the dark energy that drives the accelerating expansion of the Universe today may have also originated from the original inflationary scalar field. Newly created dark matter particles may cluster with matter through gravitational interactions, and form galaxies, with baryonic  matter clustering in its center. The initial conditions of the reheating as well as the physics of the decay of the scalar fields are not fully known, and they may need physics beyond the Standard Model of particle physics to account for the dark matter creation process. And certainly we need more accurate observational data as well, to constrain the models and find out what truly happens in the early Universe.

\section*{Acknowledgments}

We would like to thank to the three anonymous reviewers for comments and suggestions that helped us to significantly improve our manuscript. T. H. would like to thank to the Institute of Advanced Studies of the Hong Kong University of Science and Technology for the kind hospitality offered during the preparation of this work.  S.-D. L. would like to thank to the Natural Science Funding of Guangdong Province for support (2016A030313313).

\end{document}